\documentclass[journal,final]{IEEEtran}
\usepackage{amsmath,amsfonts}
\usepackage{algorithmic}
\usepackage{algorithm}
\usepackage{array}
\usepackage[caption=false,font=normalsize,labelfont=sf,textfont=sf]{subfig}
\usepackage{textcomp}
\usepackage{stfloats}
\usepackage{url}
\usepackage{verbatim}
\usepackage{graphicx}
\usepackage{cite}
\usepackage{bbding}
\usepackage{threeparttable}
\usepackage{booktabs} 
\usepackage[justification=centering]{caption}
\usepackage{makecell}
\usepackage{threeparttable}
\usepackage{multirow}
\usepackage{verbatim}

\newcommand{\Vecm}[1]{\boldsymbol{#1}}

\usepackage[dvipsnames]{xcolor}
\usepackage{subfig}

\definecolor{mygray}{gray}{0.6}

\usepackage[]{review}
\setrevision{1}

\begin{document}

\title{Self-supervised Audio Teacher-Student Transformer for Both Clip-level and Frame-level Tasks}

\author{Xian Li, Nian Shao, and Xiaofei Li$^{*}$

\thanks{\{lixian, shaonian, lixiaofei\}@westlake.edu.cn 
}
\thanks{$^{1}$Westlake Institute for Advanced Study \& $^{2} $Westlake University  , Hangzhou, China}
\thanks{* corresponding author}
}


\maketitle

\begin{abstract}
 \addnote[abstract]{1}{
Self-supervised learning (SSL) has emerged as a popular approach for learning audio representations. One goal of audio self-supervised pre-training is to transfer knowledge to downstream audio tasks, generally including clip-level and frame-level tasks.  While frame-level tasks are important for fine-grained acoustic scene/event understanding, prior studies primarily evaluate on clip-level downstream tasks.  In order to tackle both clip-level and frame-level tasks, this paper proposes Audio Teacher-Student Transformer (ATST), with a clip-level version (named ATST-Clip) and a frame-level version (named ATST-Frame), responsible for learning clip-level and frame-level representations, respectively. Both methods use a Transformer encoder and a teacher-student training scheme.  We have carefully designed the view creation strategy for ATST-Clip and ATST-Frame. Specifically, ATST-Clip uses segment-wise data augmentations, and ATST-Frame integrates frame-wise data augmentations and masking. Experimental results show that our ATST-Frame model obtains state-of-the-art (SOTA) performances on most of the clip-level and frame-level downstream tasks. Especially, it outperforms other models by a large margin on the frame-level sound event detection task. In addition, the performance can be further improved by combining the two models through knowledge distillation. Our code is available online.
 }

\end{abstract}


\section{Introduction}
 \IEEEPARstart{A}{udio} self-supervised learning (SSL), which learns knowledge from a large amount of unlabeled audio data, has emerged as a popular approach for learning audio representations  \cite{oord_representation_2019,tagliasacchi_pre-training_2020,saeed_contrastive_2020,fonseca_unsupervised_2021,niizumi_byol_2021,gong_ssast_2022,baade_mae-ast_2022,srivastava_conformer-based_2022,huang_masked_2023,chen_beats_2022}. 
 
 \addnote[siamese1]{1}{The siamese models\cite{liu_audio_2022,saeed_contrastive_2020,fonseca_unsupervised_2021,niizumi_byol_2021,niizumi_byol_2023} maximize the embedding similarity of two augmented views of the same audio clip, having shown a great promise for learning good audio representations.} Another promising technical line for audio SSL follows the spirit of BERT (Bidirectional Encoder Representations from Transformers) \cite{devlin_bert_2019}, using Transformer encoder\cite{vaswani_attention_2017} and performing a predictive task for the masked frames\cite{gong_ssast_2022,baade_mae-ast_2022,srivastava_conformer-based_2022,huang_masked_2023,chen_beats_2022}.  

 One goal of audio self-supervised pre-training is to transfer knowledge to downstream audio tasks. Generally speaking, audio tasks are defined within two different ways, i) clip-level tasks are to classify the acoustic scene or event of an entire audio clip, e.g. audio tagging, musical instrument recognition, etc., and ii) frame-level tasks are to detect and recognize event-level timestamps from an audio clip, e.g. sound event detection (SED). Previous studies primarily evaluate their methods on clip-level audio tasks, leaving the performance on frame-level audio tasks unclear. Clip-level tasks currently account for the majority of the downstream audio tasks. Only a few frame-level tasks have been well-defined in the field, such as speaker diarization and sound event detection. However, the frame-level tasks are more important for fine-grained acoustic scene/event understanding, and they are generally more challenging than clip-level tasks. To handle both clip-level and frame-level tasks, there are several issues to be considered.
 

In terms of the training criterion, a portion of previous methods focus on learning global representation of an audio clip by using clip-level training criteria\cite{saeed_contrastive_2020,niizumi_byol_2021,niizumi_byol_2023}, while others propose learning local frame-wise or patch-wise representations by using frame-level \cite{gong_ssast_2022,baade_mae-ast_2022} or patch-level criteria\cite{gong_ssast_2022,baade_mae-ast_2022,niizumi_masked_nodate,niizumi_masked_2023,huang_masked_2023,chen_beats_2022}. Most of the clip-level methods allow for extracting frame-wise representations from the intermediate output. However, since the frame-wise representations are not explicitly trained during pre-training, it is questionable whether a model trained by clip-level criterion can perform well on frame-level downstream tasks.

Besides the training criterion, a high temporal resolution for frame-level representations is necessary for frame-level tasks. For Transformer-based methods, the temporal resolution is determined by how the input sequence is organized, either patch-wisely or frame-wisely. SSAST (Self-Supervised Audio Spectrogram Transformer ) \cite{gong_ssast_2022} and MAE-AST (Masked Autoencoding Audio Spectrogram Transformer) \cite{baade_mae-ast_2022} have shown that the patch-wise strategy and frame-wise strategy perform differently for different downstream tasks. Other studies\cite{huang_masked_2023,chen_beats_2022,niizumi_masked_nodate,niizumi_masked_2023} only use the patch-wise strategy. Generally, the frame-wise strategy has a better temporal resolution than the patch-wise strategy, and thus may be more suitable for frame-level downstream tasks.  

Accounting for learning both clip-level and frame-level audio representations, this paper proposes two models: ATST-Clip and ATST-Frame, where ATST stands for Audio Teacher-Student Transformer. They are developed based on the teacher-student scheme of Bootstrap Your Own Latent (BYOL) \cite{grill_bootstrap_2020} and BYOL for Audio (BYOL-A) \cite{niizumi_byol_2021}. They both use Transformer encoder and frame-wise strategy. ATST-Clip and ATST-Frame are responsible for learning global and frame-wise representations by using a clip-level and frame-level training criterion, respectively. This work is a continuation of our previous conference paper \cite{li22p_interspeech}, in which ATST was first proposed for clip-level representation learning, which is renamed ATST-Clip in this paper to avoid ambiguity. This paper proposes a new ATST-Frame model, and a combination method of the two models based on knowledge distillation. In addition, the proposed models have been more thoroughly evaluated in this paper.

\addnote[teacher-student-networks]{1}{ATST-Clip draws inspirations from the teacher-student scheme of BYOL \cite{grill_bootstrap_2020} and BYOL-A\cite{niizumi_byol_2021}, which contains a teacher network and a student network. Give an audio clip, two different views are created through augmentations, e.g. randomly cropping at time dimension. The two views are then separately fed into the teacher network and the student network. Considering the similarity of the two views, the student network weights are updated by maximizing the embedding similarity of the two views. The teacher network weights, on the other hand, are updated by taking exponential moving average (EMA) of the student network weights.}  ATST-Clip proposes to replace the convolutional neural network (CNN) encoder of BYOL-A with a Transformer encoder, which shows a clear superiority over the CNN encoder, especially for learning the long-term semantic information of speech. More importantly, a new view creation strategy is proposed to fully leverage the capability of Transformer encoder. BYOL-A uses one short segment to create two views.  Instead, we propose to use two different long segments to create the two views, which is more fit for Transformer, as the network can learn longer temporal dependencies. The length of segments is carefully studied to control the distinction and overlap of the two segments, which is especially important for rationalizing the difficulty of matching the representations of the two views at latent space.

ATST-Frame extends ATST-Clip to explicitly learn frame-wise representations by maximizing the agreement of student's frame-level embeddings to the teacher's frame-level embeddings. Creating proper views for teacher and student branches is the key for achieving a proper difficulty for matching the frame-level embeddings, and thus guiding the model to learn meaningful frame-level representations. Both teacher and student branches process the entire audio clip to maintain the frame-to-frame correspondence between their output sequences. To increase the matching difficulty, data augmentation is applied to one of the teacher and student branches. Moreover, masking is further applied to the student branch to encourage the model to learn semantic relations between frames by accomplishing the prediction of masked frames. Our experiments show that data augmentation and masking are both necessary and are a good combination for frame-level audio pre-training within the teacher-student framework.

Finally, as the training criterion of ATST-Frame and ATST-Clip are totally different, they could learn complementary features. We propose to combine ATST-Frame and ATST-Clip at the fine-tuning stage of downstream tasks, based on cross-model knowledge distillation, which outperforms ATST-Frame or ATST-clip alone.     

We use the large-scale AudioSet \cite{gemmeke_audio_2017} for pre-training, and evaluate the models with a variety of clip-level downstream tasks and two frame-level downstream tasks. Downstream tasks cover \addnote[audiotype]{1}{multiple
 audio domains: environmental sound, speech, and music.}  Our results show that i) on clip-level tasks, after fine-tuning, the proposed models outperform other state-of-the-art (SOTA) methods for most of the tasks. Especially, the precision on the AudioSet-2M and AudioSet-20K datasets reach a new SOTA of 49.7\% and 40.5\% (without model ensembling), respectively. ii) on the frame-level SED task, the proposed ATST-Frame model performs particularly well, outperforming ATST-Clip and other methods by a large margin. \addnote[code]{1}{We open-source our code online\footnote[1]{https://github.com/Audio-WestlakeU/audiossl} for the research community to replicate and expedite future research.}

\section{Related Works}






This section introduces related works on audio self-supervised learning.

\subsection{Siamese Models}

\addnote[siamese]{1}{Siamese models use a two-tower architecture, in which each tower processes a view of the data sample and the embedding similarity of the two views are maximized during training\cite{chen_exploring_2020,liu_audio_2022}. This idea often confronts the issue of model collapse, e.g. the model can find an easy solution to output a constant value for any inputs. Various training strategies are developed to avoid model collapse. Most of these strategies are originally developed in image SSL pre-training and then are adopted by audio SSL pre-training. One of the strategies is contrastive learning, which introduces negative samples and not only pull the two views close in the latent space but also push them far away from negative samples in the latent space, e.g. SimCLR (Simple Framework for Contrastive Learning of Visual Representations)\cite{chen_simple_2020} in image SSL and its audio counterparts COLA (COntrastive Learning
 for Audio)\cite{saeed_contrastive_2020,fonseca2021unsupervised}. However, the negative samples are possibly similar to positive samples in some scenarios, which will harm the pre-training performance. Due to this reason, some recent works investigated to train siamese models without using negative samples, e.g. BYOL\cite{grill_bootstrap_2020} in image SSL pre-training and its audio counterpart BYOL-A\cite{niizumi_byol_2021}. As this work is inspired by the BYOL-style strategy, we will introduce the framework of BYOL-A in Section \ref{sec:baseline}. }


Our ATST-Clip extends BYOL-A to use Transformer encoder, and proposes a new view creation strategy to fit the Transformer encoder. Our ATST-Frame further extends ATST-Clip to explicitly learn frame-wise representations.

\subsection{Masked Audio Modelling }
Other methods follow the line of  Masked Language Modelling (MLM)\cite{devlin_bert_2019}.  This kind of method has been first applied to speech self-supervised pre-training\cite{liu_mockingjay_2020,baevski_wav2vec_2020,hsu_hubert_2021,liu_tera_2021,baevski_data2vec_2022}, and then to audio self-supervised pre-training, e.g. SSAST\cite{gong_ast_2021}, Conformer-based audio SSL method\cite{srivastava_conformer-based_2022}, MAE-AST\cite{baade_mae-ast_2022} and Audio-MAE (Audio Masked Autoencoders)\cite{huang_masked_2023}. \addnote[idea]{1}{The idea is to mask an arbitrary region of the input, and then perform a prediction task on the masked region. Some of them
train the model by reconstructing the masked region\cite{liu_mockingjay_2020,liu_tera_2021}, while others replace the reconstruction loss with a classification loss.} Wav2vec2\cite{baevski_wav2vec_2020} and its follower\cite{srivastava_conformer-based_2022}, a method based on Conformer (Convolution-augmented Transformer), solve a frame-level contrastive problem by introducing positive and negative frames. SSAST\cite{gong_ssast_2022} jointly solves a masked reconstruction and a wave2vec-style contrastive problem. HuBERT (Hidden-Unit BERT)\cite{hsu_hubert_2021} creates pseudo classification labels by performing clustering on MFCC (Mel-Frequency Cepstral Coefficient) features or output features of the model trained in the previous iteration. BEATs (Bidirectional Encoder representation from Audio Transformers)\cite{chen_beats_2022} proposes an iterative audio pre-training framework, where an acoustic tokenizer and an audio SSL model are iteratively optimized.
From the perspective of model architecture, some works \cite{baade_mae-ast_2022,10095691,niizumi_masked_nodate,huang_masked_2023} follow the asymmetric encoder-decoder structure of masked autoencoders (MAE)\cite{he2021masked}, in which the encoder encodes the unmasked region, while the decoder processes both the masked and unmasked regions and reconstructs the masked region.

The most similar works with our ATST-Frame are data2vec\cite{baevski_data2vec_2022} and M2D (Masked Modeling Duo)\cite{niizumi_masked_2023}. \addnote[match]{1}{They both use a teacher-student scheme, in which the student encodes a masked version of the training sample and the teacher provides the training/prediction target for the student. The teacher take as input either the unmasked version of the same training sample (data2vec) or only the masked parts of the student input (M2D). Besides, they both use a frame/patch-level criterion.}. However, there exist several major differences: i) data augmentation is applied in our ATST-Frame, but not in data2vec and M2D. Data augmentation is critical for adjusting the prediction difficulty; ii) \addnote[teachingrepresentation]{1}{data2vec constructs the training/prediction target for the student network by taking the average of the last eight Transformer blocks of the teacher encoder},  while our ATST-Frame uses the asymmetric structure of the BYOL\cite{grill_bootstrap_2020}, where an extra predictor network is set for the student branch. iii) M2D organizes spectrograms patch-wisely and uses a MAE structure in the student branch, while ATST-Frame adopts a frame-wise strategy and uses a regular Transformer encoder architecture.  

\section{The Proposed Method}

Two models are proposed in this work: ATST-Clip and ATST-Frame. ATST-Clip focuses on learning the global representation of an audio clip, while ATST-Frame focuses on learning frame-wise representations. Both of them use a Transformer encoder\cite{vaswani_attention_2017} to process audio spectrograms. And both of them are trained in a teacher-student scheme\cite{grill_bootstrap_2020}, in which the teacher model is updated by an exponential moving average (EMA) of the student model, while \addnote[between]{1}{the student model is updated by maximizing the similarity of its embedding to the embedding of teacher.} 

We will introduce the baseline teacher-student scheme in Section \ref{sec:baseline}, the Transformer encoder in Section \ref{sec:ast}, and then present ATST-Clip and ATST-Frame in Section \ref{sec:ATST-Clip} and Section \ref{sec:ATST-Frame}, respectively. The combination of ATST-Clip and ATST-Frame is presented in Section \ref{sec:combine}.

\subsection{Baseline Teacher-Student Scheme}
\label{sec:baseline}
\label{sec:byol-a}

In this work, we adopt the teacher-student scheme as our baseline framework, which was first proposed by Bootstrap you own latent (BYOL) \cite{grill_bootstrap_2020} for image pre-training, and adopted by BYOL-A \cite{niizumi_byol_2021} for audio pre-training. In BYOL-A, given one augmented view of an audio clip, \addnote[identical]{1}{the student network is trained to predict a data representation being close to the teacher network's representation on another augmented view of the same audio clip. During training, the teacher network weights are updated by taking the EMA of the student network weights. }

Formally, the student network, defined by a set of weights $\theta$, contains an encoder $f_{\theta}$, a projector $g_{\theta}$ and a predictor $q_{\theta}$, while the teacher network, defined by a set of weights $\phi$, contains only an encoder $f_{\phi}$ and a projector $g_{\phi}$. \addnote[encoder-projector-predictor]{1}{The encoder, a CNN in BYOL and BYOL-A, extracts representations from the augmented views. The projectors and the predictor are multi-layer perceptrons (MLPs) that consist of a linear layer (with output dimension of 4096) followed by batch normalization, rectified linear units (RELU), and a final linear layer (with output dimension of 256). The output representation of encoder is used in downstream tasks. Using projectors in pre-training is shown to improve the representation quality \cite{chen_simple_2020}. It has been shown that the additional predictor in the student network (combined with the stop-gradient operation of teacher network) is the key factor for preventing the model from collapsing \cite{chen_exploring_2020}.} During training, $\phi$ is updated by the EMA of $\theta$ as: $\phi \leftarrow m\phi + (1-m)\theta$, where $m$ is a decay rate. $\theta$ is updated as follows. Let $(\Vecm{X},\Vecm{X}' )$ be two views created from an audio clip.  $\Vecm{X}$ is fed into the teacher network to obtain $\Vecm{h}=f_{\phi}(\Vecm{X})$ and $\Vecm{z}=g_{\phi}(\Vecm{h})$. $\Vecm{X'}$ is fed into the student network to obtain $\Vecm{h'}=f_{\theta}(\Vecm{X}')$, $\Vecm{z}'=g_{\theta}(\Vecm{h}')$ and $q_{\theta}(\Vecm{z}')$. $\Vecm{z}$ and $q_{\theta}(\Vecm{z}')$ are then L2-norm normalized to  $\overline{\Vecm{z}}$ and $\overline{q}_{\theta}(\Vecm{z'})$, and the mean square error (MSE) loss between them is computed:
\begin{equation}
    L_{\theta}= \Vert \overline{\Vecm{z}}-\overline{q}_{\theta}(\Vecm{z'}) \Vert_2^2
    \end{equation}
 A symmetric loss $L_{\theta}'$ is also calculated by feeding $\Vecm{X}$ to the student network and $\Vecm{X}'$ to the teacher network. During training, $\theta$ is updated by minimizing $L_{\theta}^{total} = L_{\theta} + L_{\theta}' $. 
 
In BYOL-A, the encoder is a CNN.The proposed models will replace the CNN encoder with a Transformer encoder, and use the same projectors and predictors as BYOL-A.

\subsection{Audio Spectrogram Transformer Encoder}
\label{sec:ast}

Both ATST-Clip and ATST-Frame use the same encoding network architecture. The raw waveform is first converted to log-mel spectrogram $\Vecm{X} \in \mathbb{R}^{L\times C}$, where $L$ and $C$ denote the number of frames and the number of frequency bins, respectively. Since modeling long sequences with Transformer is computationally demanding, four consecutive frames of $\Vecm{X}$ are stacked as one frame to reduce the sequence length. The stacked frames are fed to a linear projection layer (with output dimension of $d$) to obtain a new embedding sequence $\Vecm{E}\in \mathbb{R}^{\frac{L}{4}\times d}$ as the input sequence of Transformer encoder. 
The embedding sequence is then added with a trainable absolute lookup table positional embedding $\Vecm{P} \in \mathbb{R}^{(\frac{L}{4}) \times d} $. 
Eventually, the embedding sequence is processed by a Transformer encoder, obtaining an output embedding sequence of $\Vecm{O} \in \mathbb{R}^{(\frac{L}{4})\times d}$. \addnote[pre-ln]{1}{Our Transformer encoder architecture is the same as Vision Transformer \cite{vits}, which is a Pre-LN (Layer Norm) Transformer\cite{pre-ln}.}  To represent the entire clip, ATST-Clip incorporates an extra trainable class token $\texttt{[CLS]} \in \mathbb{R}^{1\times d}$, which will be detailed in \ref{sec:ATST-Clip}.

\begin{figure*}[!t]
\centering
\subfloat[ATST-Clip]{\includegraphics[width=2.5in]{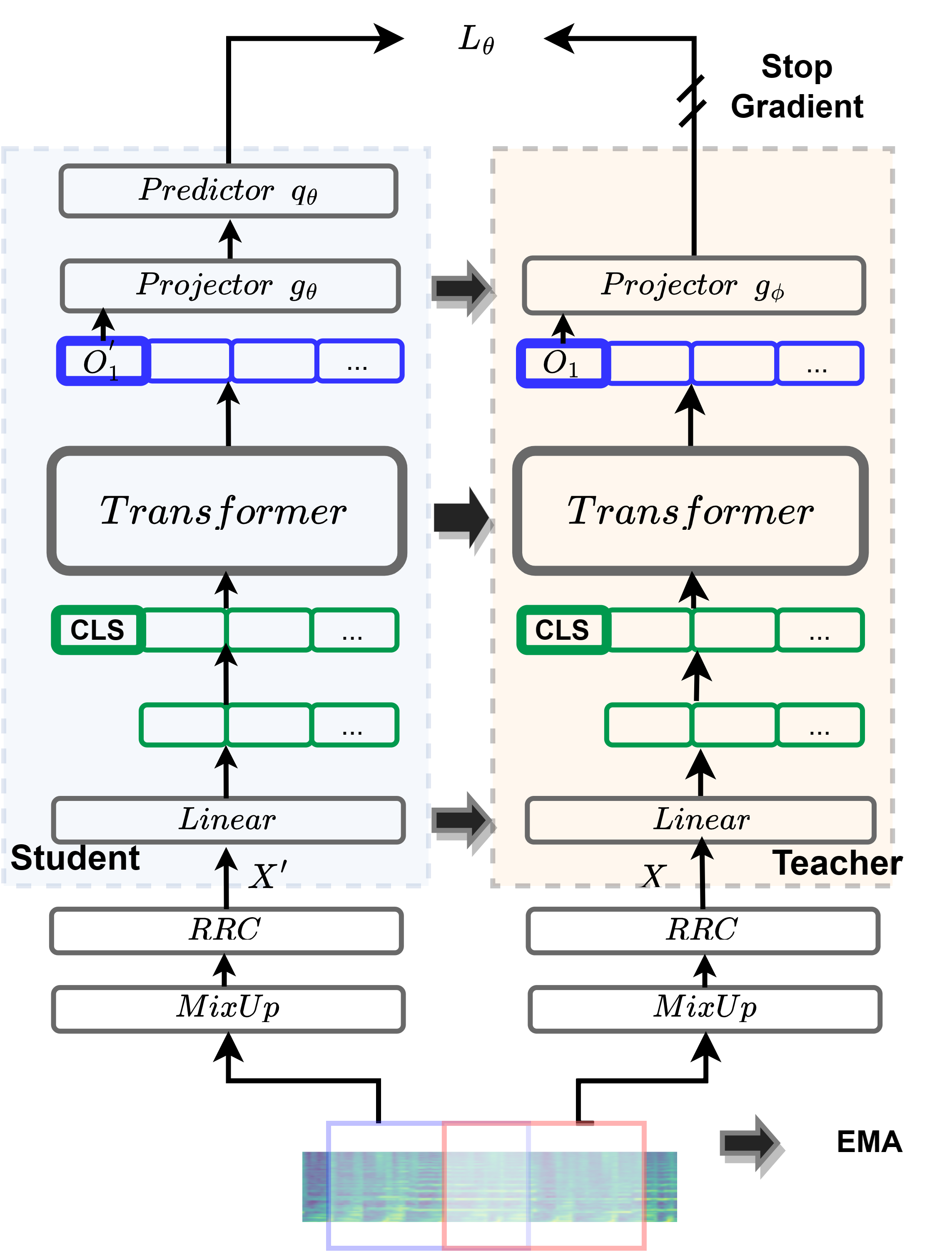}}
\hfil
\subfloat[ATST-Frame]{\includegraphics[width=2.5in]{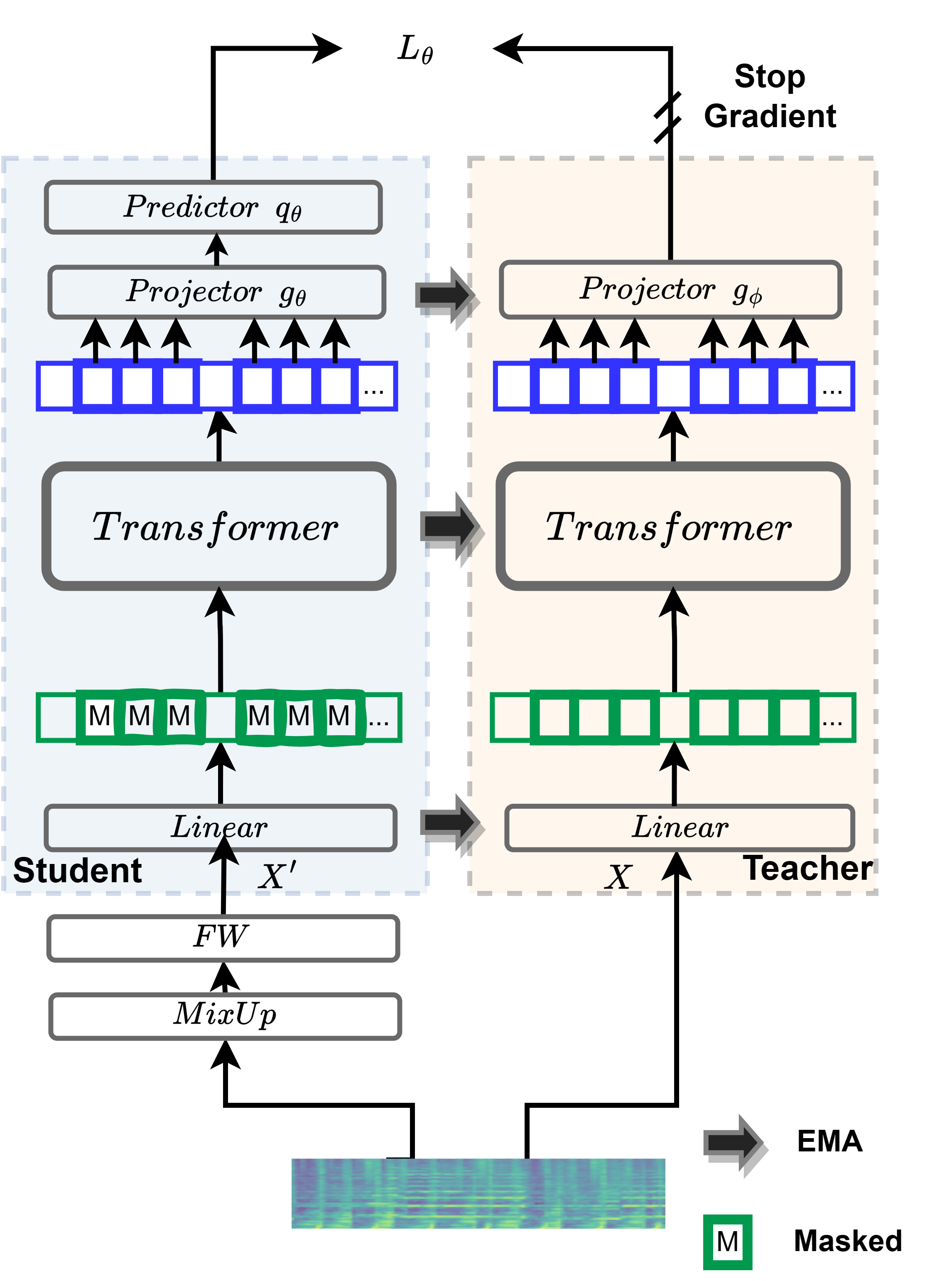}}%

\caption{ The proposed methods. (a) ATST-Clip (b) ATST-Frame. The loss $L_{\theta}$ is computed by feeding $X$ to the teacher branch and $X'$ to the student branch. The symmetric loss $L_{\theta}'$ can be computed by swapping $X$ and $X'$ (not shown in the figure).}
\label{fig:framework}
\end{figure*}

\subsection{ATST-Clip}
\label{sec:ATST-Clip}

The major difference between ATST-Clip and BYOL-A is twofold. ATST-Clip uses a Transformer  encoder to leverage its powerful abilities in modeling long-term dependencies and uses a new view creation strategy specifically fit for the Transformer encoder.

\addnote[transformation]{1}{First, given an audio clip, it is converted from the waveform domain to the log-mel spectrogram, from which two views are created through a set of augmentations.} The two views are then fed into the student and teacher branches respectively, generating a clip-level representation at each branch. In the end, the two clip-level representations are used to calculate the training loss.

\subsubsection{Creation of Views}

\addnote[contrastive2siamese]
{1}{For siamese model, the two views should be similar with each other to be identified as the same sample yet different enough to increase the difficulty of the identification}. BYOL-A \cite{niizumi_byol_2021} randomly crops a single 1-second segment from the input audio and then creates two views by applying different data augmentations to this single segment. It is considered in BYOL-A \cite{niizumi_byol_2021} that different segments may be too different to be identified \addnote[positivepair]{1}{as the same sample}. The work in \cite{fonseca_unsupervised_2021} uses two segments to create two views, however, it uses negative samples to mitigate the problem caused by using two segments.   

Our view creation strategy is shown in Fig.~\ref{fig:framework}(a). 
The time domain input audio clip is first transformed to log-mel spectrogram. We randomly crop two different segments from the log-mel spectrogram. Then, two types of data augmentation are applied to each of the segments, creating two views of the input audio clip, i.e. $(\Vecm{X},\Vecm{X}')$. The augmentations we employed include Mixup\cite{niizumi_byol_2021} (a modified version of the original Mixup\cite{bcsound,mixup}) and Random Resize Cropping (RRC)\cite{niizumi_byol_2021} (adapted from RRC\cite{szegedy2015going} in computer vision to accommodate audio signals).   

In order to take full advantage of the Transformer's ability in modeling long-term dependencies, the proposed method intends to use longer segments, e.g. \addnote[6seconds]{1}{6-second segments randomly cropped from 10-second training audio clips in our experiments.} The proposed method separately creates two views from two different segments for the purpose of increasing the difficulty of identifying the two views as the same sample, thus leading the model to learn more generalized representations. On the other hand, the two segments cannot be too far away from each other, otherwise, the similarity between them is completely lost. This is guaranteed by properly setting the segment length to make the two segments have a certain portion of overlap. 
Overall, the proposed strategy does not lose the rationality of identifying two segments as the same sample due to the overlap constraint, and meanwhile increases the task difficulty by using two segments and thus helping to learn more generalized presentation. 

\subsubsection{Encoding}
The encoding procedure is illustrated in Fig.~\ref{fig:framework}(a). To obtain a representation for the entire clip, an extra class token is used. \addnote[afterlinear]{1}{First, a linear projection layer processes the view, $X$ or $X'$, obtaining an embedding sequence, at the beginning of which a trainable class token  $ \texttt{[CLS]}\in \mathbb{R}^{1\times d}$ is inserted.} The embedding sequence is then added with \addnote[pos_emb]{1}{a trainable absolute lookup table positional embedding sequence, and then fed into the encoder.} The class token $ \texttt{[CLS]}$ is widely used for sentence embedding in neural language processing \cite{devlin_bert_2019}, global image embedding \cite{caron_emerging_2021}, as well as audio segment embedding \cite{gong_ast_2021}. It aggregates information from the embedding sequence at every Transformer blocks with the self-attention mechanism. 
In the output embedding sequence, the class token, denoted as $\Vecm{O_1}\in \mathbb{R}^{1 \times d}$, is taken as the final clip representation. $\Vecm{O_1}$ is then processed by the following projector (and predictor). 

\subsubsection{Loss Function}

\addnote[loss_clip]{1}{The loss function is the same as the one in the baseline scheme described in Section \ref{sec:byol-a}. }

\subsection{ATST-Frame}
\label{sec:ATST-Frame}

As BYOL-A and ATST-Clip have shown powerful abilities in learning clip-level audio representations with the teacher-student scheme, we further adopt the teacher-student scheme to develop ATST-Frame, which explicitly learns fine-grained frame-wise representation. ATST-Clip creates two different views of the audio clip, and then maximizes the agreement between the clip-level representation of the two views. Instead, ATST-Frame maximizes the agreement between the frame-level representations of two views. The key is to properly design the two views to achieve a good trade off between the difficulty and rationality of the frame-level pretext task. 

\subsubsection{Creation of Views}
Different from ATST-Clip which randomly crops the audio clip, ATST-Frame processes the entire audio clip. The reasons are i) the frame correspondence of the two views should be preserved for measuring the frame-level agreement; and ii) in order to take full advantage of  Transformer in modeling long-term dependencies, the views are set to be as long as possible.

To increase the difference of the two views, and thus increase the task difficulty, data augmentation is first applied to one of the two views. This time, the augmentations should preserve the frame correspondence of the two views. Two augmentations are used: Mixup \cite{niizumi_byol_2021} and Frequency Warping (FW). For computational efficiency, FW is implemented in the spectrogram domain through cropping and then resizing at the frequency axis. Specifically, the input log-mel spectrogram $\Vecm{X} \in \mathbb{R}^{L \times C }$, is first cropped at the frequency axis as $\Vecm{X}_{1:L,1:a}$, and then resized by bi-cubic interpolation at the frequency axis as $\Vecm{X}^{FW} \in \mathbb{R}^{L \times C}$, where $C$ is the number of the mel frequency bins, and the integer number $a$ is uniformly sampled from the frequency range of $[C*0.6, C ]$. These operations lead to an approximate yet efficient frequency warping. 

Due to the constraint of frame-to-frame correspondence for the two views, data augmentation does not bring sufficient task difficulty. Thence, we adopt \addnote[bert-masking]{1}{BERT-like masking \cite{devlin_bert_2019}, which masks/replaces a portion of the frames with a certain trainable mask token and then performs a prediction task to predict the masked frames.} The student network takes as input a masked version of the training sample and learns to predict the masked frames, while the prediction target is provided by the teacher network taking as input the unmasked version of the training sample. 
To prevent the model cheating by simply interpolating, we adopt the group masking strategy \cite{baevski_wav2vec_2020} that forces $N$ adjacent frames to be masked together. Specifically, we set a probability of 0.65 for masking and force five adjacent frames to be masked together as a masked block, and the masked blocks are allowed to overlap. With this setting, approximately 50\% of the frames are masked.  As will be explained later, the pre-training loss will be computed only on the masked frames. 

Overall, combining data augmentation and masking is able to create two proper views for the frame-wise pretext task within the teacher-student framework. Although these techniques, i.e. data augmentation, masking and teacher-student scheme, have already been individually (or together with other techniques) used for audio pre-training in the literature, this work carefully integrates them in a different way from other methods, and achieves noticeably better performance. 
For example, frame-level training of other teacher-student schemes 
\cite{baevski_data2vec_2022,atito_asit_2022,niizumi_masked_2023} do not use data augmentation. And other masking-based methods \cite{baade_mae-ast_2022,huang_masked_2023} reconstruct the masked region.


\subsubsection{Encoding}
 The encoding procedure is illustrated in Fig.~\ref{fig:framework}(b). The input log-mel spectrogram is first data augmented with Mixup and FW for one view, then processed with a linear projection.  After the linear projection, the embedding sequence $\Vecm{E}\in \mathbb{R}^{\frac{L}{4}\times d}$ is randomly masked along the time dimension, only for the student branch. Each masked frame is substituted with a trainable vector $\Vecm{M}\in \mathbb{R}^{1\times d}$. Subsequently, the embedding sequence is added with \addnote[pos_emb2]{1}{a trainable absolute lookup table positional embedding sequence, and then fed into the encoder.} After the encoder, the unmasked frames are thrown away, and only the masked frames are further processed by the following projector (and predictor). 

\subsubsection{Loss Function}

 \addnote[loss_frame]{1}{The loss function of ATST-Frame differs from the one of baseline or ATST-Clip in the sense that the loss of ATST-Frame is computed frame-wisely. Feeding the two views, i.e. $\Vecm{X}$ and $\Vecm{X}'$, to the teacher branch and the student branch,  we obtain $\Vecm{z} \in \mathbb{R}^{N_{\text{mask}} \times 256}$ and $q_{\theta}({\Vecm{z'}})\in \mathbb{R}^{N_{\text{mask}} \times 256}$, respectively, where $N_{\text{mask}}$ denotes the number of masked (stacked-)frames}. The MSE loss is calculated on the L2-normalized embeddings $\overline{\Vecm{z}}$ and $\overline{q}_{\theta}(\Vecm{z'})$ as
\begin{equation}
    L_{\theta} = \frac{1}{N_{\text{mask}}} \sum_{i=1}^{N_{\text{mask}}} \Vert  \overline{\Vecm{z}}_i-\overline{q}_{\theta}(\Vecm{z'}_i)  \Vert_2^2.
\end{equation}
A symmetric loss $L_{\theta}'$ is also calculated by feeding $\Vecm{X}$ to the student network and $\Vecm{X}'$ to the teacher network. During training, $\theta$ is updated by minimizing $L_{\theta}^{total} = L_{\theta} + L_{\theta}' $.

\subsection{Combine ATST-Clip and ATST-Frame}
\label{sec:combine}

ATST-Clip and ATST-Frame focus on learning global clip-level representation and
local frame-level representations, respectively. Combining them may yield more comprehensive representations. ATST-Clip is trained by a clip-level criterion, but still can extract frame-wise representations with the encoder. However, these representations are only used for storing local information from which the class token $ \texttt{[CLS]}$ can aggregate information, but are not optimized specifically to fit frame-level downstream tasks. On the other hand, ATST-Frame can obtain a clip-level representation by applying average pooling to the frame-wise representations. However, this type of average information is not optimized specifically to fit clip-level downstream tasks.

It is possible to jointly train ATST-Clip and ATST-Frame, e.g. add a \texttt{[CLS]} token to ATST-Frame and then perform multi-task learning. Actually, ASiT \cite{atito_asit_2022} is trained with a combination of three tasks, including a global task of maximizing agreement, a local task of maximizing agreement, and a reconstruction task. 
However, our preliminary experiments show that, when creating two views, it is hard to trade off the task difficulty for both ATST-Clip and ATST-Frame. ATST-Clip requires two different randomly cropped segments and the two segments have only a certain portion of overlap, while ATST-Frame asks for a frame-to-frame correspondence between the two views. Besides, ATST-Clip and other clip-level audio siamese methods\cite{fonseca2021unsupervised,niizumi_byol_2021,niizumi_byol_2023} largely leverage the RRC augmentation \cite{niizumi_byol_2021} to achieve a good performance, but RRC will distort the frame-to-frame correspondence.

Therefore, we leave ATST-Clip and ATST-Frame trained separately to maintain their own advantages and combine them in the evaluation stage. A straightforward way is to ensemble the two models at the inference stage, \addnote[straightforward]{1}{e.g. to concatenate the outputs of the two models}, but this will double the computational cost for inference. 
Instead, we use knowledge distillation \cite{hinton2015distilling} to combine the two models.  CMKD \cite{gong_cmkd_2022} has explored cross-model knowledge distillation in the context of supervised audio tagging, e.g. distilling knowledge from EfficientNet-B0 \cite{gong_psla_2021} to AST \cite{gong_ast_2021}. We follow the principle of CMKD. We first fine-tune ATST-Clip on a downstream task, and then use the fine-tuned ATST-Clip as a teacher to fine-tune ATST-Frame on the same downstream task. \addnote[cmkd_specify]{1}{Specifically, ATST-Frame is fine-tuned by using two classification losses computed with the ground-truth labels and the ATST-Clip predictions, respectively, and the two losses are weighted with a balance term $\lambda=0.5$. }
This strategy is denoted as ATST-C2F. Or we can reverse the fine-tuning order to have the ATST-F2C strategy. 
 These strategies approximately double the fine-tuning time compared with ATST-Frame or ATST-Clip alone, but do not increase the computational cost for inference. 

 \subsection{Transferring to Downstream Task}
\label{sec:transfer-downstream}
For both ATST-Clip and ATST-Frame, after pre-training, we discard the projector in the teacher network, and use the teacher encoder to extract embeddings for downstream tasks.

\section{Experimental Setup}
\label{setup}
We conduct extensive experiments using the large-scale AudioSet\cite{gemmeke_audio_2017} for pre-training and a variety of downstream tasks for evaluation.  The evaluation is performed under the protocol of linear evaluation or fine-tuning.  In linear evaluation, the pre-trained encoder is frozen as a feature extractor, on top of which a linear classifier is trained. Whereas in fine-tuning, the pre-trained encoder and linear classifier are fine-tuned together.

\begin{table}[t]
\centering
\begin{threeparttable}
\begin{tabular}{l|ccccc}
\toprule
~ &  \#parameters &  \#blocks &  \#heads &  dimension &   \\
\midrule
ATST-Clip$_{small}$ & 22M & 12 & 6  & 384  \\

ATST-Clip  & 86M & 12 & 12 & 768 \\

ATST-Frame$_{small}$ & 22M & 12 & 6  & 384  \\

ATST-Frame  & 86M & 12 & 12 & 768 \\
\bottomrule
\end{tabular}
\end{threeparttable}

\caption{The size of models. }
\label{tab:detail_model}
\end{table}

\subsection{Pre-training}

We use AudioSet \cite{gemmeke_audio_2017} for pre-training. The full AudioSet contains 2 million audio clips captured from Youtube videos, with a fixed clip length of 10 seconds. \addnote[unbalancedset]{1}{The AudioSet is published with an unbalanced set with 2,042,985 clips and a balanced set with 22,176 clips. We use the unbalanced set of the AudioSet for pre-training. Due to the change of YouTube video availability, the unbalanced set we use contains 1,912,024 clips (AS-1.9M). }

For both ATST-Clip and ATST-Frame, a base model is trained using AS-1.9M, which contains 12 Transformer encoder blocks, and 12 heads for each block. The dimension and inner dimension are 768 and 3072, respectively. Besides, we also trained a small model for them for accelerating the development process and conducting ablation studies, using a subset of 200 thousand randomly sampled audio clips (AS-200K). The small model contains 12 Transformer encoder blocks, and 6 heads for each block. The dimension and inner dimension are 384 and 1536, respectively. In the following, we use ATST-Clip and ATST-Frame to represent the base models by default, while ATST-Clip$_{small}$ and ATST-Frame$_{small}$ for the small models. 

Audio is re-sampled to 16 kHz. Audio clips are transformed to the log-mel spectrogram domain, with a Hamming window, a window length of 64 ms, a hop size of 10 ms, and 64 mel-frequency bins ranging from 60 Hz to 7800 Hz. The mel-spectrogram feature is min-max normalized, where the minimum and maximum values are calculated globally on the pre-training dataset.

\textbf{ATST-Clip:}
\label{sec:impdetail_pre}
We intentionally set the length of two segments (for creating two views) to 6 seconds, which will lead to a segment overlap of at least 2 second, considering that the length of audio clip is 10 seconds. The two randomly sampled segments are augmented by Mixup and RRC with the same configurations used in BYOL-A \cite{niizumi_byol_2021}. 
 
\textbf{ATST-Frame:}
We use the entire audio clip (10 seconds in AudioSet) for training. We set a probability of 0.65 for masking, and force five adjacent frames to be masked.  

Hyper-parameters for pre-training are listed in Table \ref{tab:hyper-pretrain}. We pre-train our models with the AdamW optimizer \cite{loshchilov2017fixing}. The learning rate is warmed up for 10 epochs, and then annealed to $10^{-6}$ at cosine rate\cite{loshchilov2016sgdr}. Similar to DINO \cite{caron_emerging_2021}, the weight decay of Transformer is increased from 0.04 to 0.4 at cosine rate. The EMA decay rate increases from an initial value to 1 at cosine rate. 

\begin{table}[t]
\centering
\scalebox{0.85}{
\begin{threeparttable}
\begin{tabular}{l|cccc}
\toprule
~ & ATST-Clip$_{small}$ & ATST-Clip & ATST-Frame$_{small}$ & ATST-Frame \\
\midrule
\makecell{Dataset} & AS-200K & AS-1.9M & AS-200K & AS-1.9M \\
\midrule
\makecell{Optimizer} & AdamW & AdamW & AdamW & AdamW \\
\midrule
\makecell{Batch\\ size} & 1536 & 1536 & 1024  & 864 \\
\midrule
\makecell{Learning\\ rate} & 5e-4 & 2e-4 & 4e-4 & 8e-5 \\
\midrule
\makecell{Warm up\\ (epochs)} & 10 & 10 & 10 & 10  \\
\midrule
\makecell{Epochs}           & 300 & 200 & 300 & 200 \\
\midrule
\makecell{Initial EMA \\Decay Rate}   & 0.99 & 0.9995 & 0.997 & 0.9996   \\
\midrule
\makecell{Initial \\Weight Decay}     & 0.04  & 0.04  & 0.04  & 0.04 \\
\midrule
\makecell{Final \\ Weight Decay}     &  0.4 &  0.4 &  0.4 & 0.4\\
\midrule
\makecell{Drop \\ path} & 0.1 &0.1 & 0.1 &0.1\\
\midrule
\makecell{Dropout} & 0 &0 & 0 &0\\
\bottomrule
\end{tabular}
\end{threeparttable}
}

\caption{Hyper-parameters for pre-training.}

\label{tab:hyper-pretrain}
\end{table}

\subsection{Clip-level Downstream Tasks}
\label{sec:dataset}

\subsubsection{Datasets} Evaluations are carried out on a variety of clip-level downstream tasks, which cover \addnote[audiotype2]{1}{multiple audio domains: environmental sound, speech and music.}

\begin{itemize}
\item \textbf{AS-20K} for multi-label sound event classification. We use the balanced set of AudioSet, with 527 audio classes. We successfully downloaded 20,886 audio clips for training and 18,886 audio clips for evaluation.
\item \textbf{AS-2M}  for multi-label sound event classification. We use the unbalanced set and balanced set (1,932,110 clips in total) together for training, and use the 18,886 audio clips for evaluation.
\item \textbf{US8K} for single-label audio scene classification. We use the Urbansound8k dataset \cite{salamon2014dataset} to classify audio clips (less than 4 seconds) into 10 classes. It contains 8,732 audio clips and has ten folds for cross-validation.
\item \textbf{SPCV2} for spoken command recognition. We use Speech Command V2 \cite{warden2018speech} to recognize 35 spoken commands for one second of audio. It contains 84,843, 9,981 and 11,005 audio clips for training, validation and evaluation respectively.
\item \textbf{VOX1} for speaker identification. We use the Voxceleb1 dataset \cite{nagrani2017voxceleb}, with 1,251 speakers. It contains 13,8361, 6,904 and 8,251 for training, validation and evaluation, respectively.
\item \textbf{NSYNTH} for musical instrument classification. We use the NSYNTH dataset \cite{engel2017neural}, to recognize 11 musical instrument family classes from 4-second audio clips.
\item \textbf{FSD50K} for multi-label sound event classification. We use FSD50K dataset\cite{fonseca_fsd50k_2022}, which contains 36,796, 4,170 and 10,231 audio clips for training, validation and evaluation, respectively.
\end{itemize}

\subsubsection{Metric} We take classification accuracy (Acc) as the performance metric for the single-label tasks, including audio scene classification, spoken command recognition, speaker identification and musical instrument classification, and mean average precision (mAP) for the tasks of multi-label sound event classification.  

For datasets containing validation set, we use the validation set for hyper-parameters tuning and model selection, and report metric score of the selected model on evluaiton set. As for AS-20K and AS-2M, we tune hyper-parameters and report metric score on evaluation set. Note this is a common practice for AudioSet fine-tuning\cite{gong_ast_2021,chen_beats_2022,huang_masked_2023}, mainly because it is non-trivial to sample a meaningful validation set from AuioSet due to extreme class imbalance and label co-occurrence. For US8K, we conduct 10-fold cross-validation, and report the average accuracy of the 10 folds. 

\subsubsection{Clip-level Embedding Extraction} 
\label{clip-level-extract}
For clip-level downstream tasks, the pre-trained models need to provide a clip-level representation. ATST-Clip is directly designed for this purpose. Although ATST-Frame is designed for frame-wise learning, the average of frame-level representations could also be a reasonable clip-level representation. Therefore, we evaluate both ATST-Clip and ATST-Frame for the clip-level downstream tasks. 

In linear evaluation experiments, we use the output of all 12 encoder blocks to construct the clip-level representation. For ATST-Clip, the embedding for the class token and the average of the rest of embedding sequence are first concatenated for each block. We find that the latter still provides some extra information besides the former. Then, the embeddings are concatenated over blocks. For ATST-Frame, the average of the embedding sequence for all blocks are concatenated. \addnote[dimension]{1}{The embedding dimensions for ATST-Clip$_{small}$, ATST-Clip, ATST-Frame$_{small}$ and ATST-Frame are $384\times 12\times2$, $768 \times 12 \times 2$, $384\times 12$ and $768\times 12$, respectively.}

In fine-tuning experiments, we only use the output of the last block. \addnote[dimension2]{1}{The embedding dimensions for ATST-Clip$_{small}$, ATST-Clip, ATST-Frame$_{small}$ and ATST-Frame are $384\times2$, $768 \times 2$, $384$ and $768$, respectively.}

The long audio clips will be split into chunks without overlap, with a chunk length of 6 seconds and 10 seconds for ATST-Clip and ATST-Frame respectively. In pre-training, ATST-Clip (ATST-Frame) processes audio clips with a fixed length of 6 seconds (10 seconds), and accordingly the length of positional embedding sequence is also 6 seconds (10 seconds). \addnote[6s10s]{1}{To account for the length of positional embedding, ATST-Clip (ATST-Frame) will also process audio clips not longer than 6 seconds (10 seconds) for downstream tasks.} Note that, audio clips that are longer than 12 seconds are first centrally cropped with a maximum length of 12 seconds, thus there will be at most two chunks for one clip. The chunks are independently processed by the pre-trained models, and their outputs are averaged to obtain the final clip representation. 

\subsubsection{Downstream Task Training}

In linear evaluation experiments, we train the linear classifier for 100 epochs with the SGD optimizer. The learning rate is annealed to $10^{-6}$ at cosine rate during training. The optimal initial learning rate is searched for each task separately. Batch size is set to 1024. Data augmentation is not used.

In fine-tuning experiments, we fine-tune all models with the SGD optimizer.  The learning rate is warmed up for 5 epochs, and then annealed to $10^{-6}$ at cosine rate\cite{loshchilov2016sgdr}. \addnote[layer-wise]{1}{The learning rate is also scheduled by a layer-wise learning rate schedule\cite{bao_beit_2022}, in which the learning rate is multiplied with a scaling factor computed with the scaling function $s(i)=\alpha^{n-i}$, where $n$ and $i$ are the number of layers and the layer index, respectively, and $\alpha$ is usually less than 1 and set to be 0.75 in our experiments.} The optimal learning rate is searched for each task separately. Batch size is set to 512 for SPCV2, Vox1, AS-20K, FSD50K, NSYNTH and 1024 for AS-2M. We trained AS-2M for 10 epochs, SPCV2, VOX1 and NSYNTH for 50 epochs, FSD50K for 100 epochs and AS-20K for 200 epochs. For SPCV2, FSD50K, NSYNTH, AS-20K and AS-2M, we use Mixup\cite{bcsound,mixup}  and RRC for data augmentation. For VOX1, data augmentation is not applied. For AS-2M, we use balance sampling\cite{gong_psla_2021}, which is a common strategy to train AudioSet, due to its unbalanced distribution of classes. Note that, the Mixup methods used for pre-training and downstream tasks are different, the former only mixes the audio clips since labels are not involved in training, while the latter mixes both audio clips and labels.

\subsection{Frame-level Downstream Task} 
\label{sec:sedsetup}

\subsubsection{Dataset}
\addnote[frame_level_task]{1}{
The evaluations of frame-level downstream tasks are conducted on the sound event detection (SED) task. SED is a frame-level multi-class classification task, which requires the model to recognize the sound events as well as their corresponding timestamps from the given audio clips. Two datasets are used for evaluation: domestic environment sound event detection (DESED) \cite{DESED} and strongly-labeled AudioSet \cite{audioset_strong}. 
}

\begin{itemize}
    \item \addnote[DESED]{1}{
        \textbf{DESED} dataset is provided by the Detection and Classification of Acoustic Scenes and Events (DCASE) Challenge 2022, task 4 - Sound Event Detection in Domestic Environment. The DESED dataset provides both labeled and unlabeled recordings for training. Since our goal is to evaluate the SSL methods, we only utilize the labeled audio clips in our experiments for linear evaluation and fine-tuning of the pre-trained models. Specifically, 1,476 clip-level labeled (weakly-labeled) real audio clips and 12,500 frame-level labeled (strongly-labeled) synthetic audio clips are used for training and validation. As for evaluation, the DCASE task 4 development set is used, containing 1,168 frame-level labeled real audio clips. It is worth mentioning that, in DCASE task 4, there are a large amount of unlabelled data used for semi-supervised training, which however are not used in this work. 
        }
    \item \addnote[AudioSetStrong1]{1}{
        \textbf{Strongly-labeled AudioSet} is a subset of the AS-2M dataset. It contains 102,561 and 15,958 frame-level labeled real audio clips for training and   evaluation, respectively. There are 407 audio classes in total. The audio classes are seriously unbalanced, where the most frequently appeared 10 event classes occupy 50.7\% of the total amount of events. 
        }
\end{itemize}
 

\subsubsection{Metric}
\label{sec:frame_metric}
The evaluation on the DESED dataset is accomplished by the official metrics of DCASE Challenge 2022, i.e. the polyphonic sound event detection scores (PSDS) \cite{MetricPSDS}. This metric measures the intersection between truth events and detected events. Two different sets of PSDS parameters are used, denoted as $\text{PSDS}_1$ and $\text{PSDS}_2$, to emphasize the low reaction time (accurate localization of sound event) and the low confusion rate between classes, respectively. For both metrics, the higher the better. \par
\addnote[AudioSetStrong2]{1}{
    The evaluation methods used in the original work of strongly-labeled AudioSet \cite{audioset_strong} have a coarse temporal resolution of 960 ms. We think they are not fine-grained enough to represent the detection accuracy, considering that the length of sound events could be as small as tens of milliseconds. 
    In our experiments, the two PSDSs used for the DESED task are used for the strongly-labeled AudioSet as well. 
    The vanilla PSDS \cite{MetricPSDS} includes an optional penalty term by the performance variance across all classes. Such penalty term evaluates the stability of performance across classes. However, the number of classes of strongly-labeled AudioSet is very high (407) and the classes are heavily unbalanced, such that all the test models in our experiments have a high performance variance across classes. With the penalty term, $\text{PSDS}$ scores could be reduced to 0 for many models, which cannot conduct fair comparison. Therefore, we will report the scores without applying the penalty term as well. 
    }


\subsubsection{Frame-level Embedding Extraction}
\label{frame-level-extract}
For this frame-level downstream task, both ATST-Clip and ATST-Frame only use the frame-level embedding sequence. In both linear evaluation and fine-tuning experiments, the embedding sequence of the last encoder block is used.
 
\addnote[6s10s_2]{1}{The length of all the audio clips in both DCASE and strongly-labeled AudioSet datasets is 10 seconds, which is exactly the same as the length of the positional embedding adopted by ATST-Frame, therefore, ATST-Frame processes these audio clips without splitting. For ATST-Clip, to account for the length of positional embedding, the audio clips are splitted into two chunks and the representations of the two chunks are concatenated along the time dimension. } 

\subsubsection{Downstream Task Training}
\label{sec:DESEDTraining}

 On top of the frame-level embedding, a multi-class classifier is added. \addnote[DESEDTraining1]{1}{For the DESED dataset, to account for the weak labels, the frame-level detection results are pooled with a softmax attention linear classifier, following the principle of DCASE challenge baseline method  \cite{softmax_attention}. Note that, our setup is different from the one of the DCASE baseline, as the latter uses some extra networks besides the pre-trained model and uses some unlabelled data. 
 } \addnote[AudioSetStrong3]{1}{As for the strongly-labeled AudioSet, a simple linear classifier is cascaded behind the frame-level embeddings to generate detection results. \\
 \indent In fine-tuning experiments, considering the limited data size of the DESED dataset, we only unfreeze the last encoder block to avoid over-fitting. The batch size is set to [128, 128] for weakly- and strongly-labeled samples, respectively. For the strongly-labeled AudioSet, we unfreeze the entire model, where the batch size is set to 256.} The models are trained with the SGD optimizer for 100 epochs. The learning rate is warmed up for 5 epochs.

\section{Results}
\subsection{Ablation Study}
Ablation experiments are conducted using ATST-Clip$_{small}$  and ATST-Frame$_{small}$ with the linear evaluation protocol, due to their low computational cost.

\subsubsection{Ablations on ATST-Clip}
\label{sec:ablation}

We separately evaluate the effectiveness of the Transformer encoder and the proposed view creation strategy. Table~\ref{tab:segments} shows the results. The result of BYOL-A is also given, which uses a CNN encoder and a single 1-second segment. Our models use single or two segments, with a length of 1 second or 6 seconds. For a fair comparison, when the segment length is set to 1 second, we split audio clips into 1-second chunks for downstream tasks.

\textbf{Transformer Encoder:} With the same view creation strategy, i.e. creating two views from a 1-second segment, our model (line 2 in Table \ref{tab:segments}) outperforms BYOL-A, especially for the two speech tasks (SPCV2 and VOX1). Speech involves more long-term semantic information, and Transformer is more suitable than CNN for learning these long-term dependencies. 

\textbf{View Creation Strategy:} As shown in Table \ref{tab:segments}, when the segment length is set to 1 second, using one single segment is better than using two segments. This phenomenon is consistent with the claim made in BYOL-A \cite{niizumi_byol_2021} that the two segments may be too different to be identified as the same sample. However, two views created from a single segment may share too much semantic content, thus leading our model to find an easy solution.
When the segment length is increased to 6 seconds, the performance measures of AS-20K, VOX1 and US8K are systematically increased, no matter whether using one or two segments. This is partially due to the capability of learning long-term dependencies of the Transformer encoder.
In addition, for the 6-second case, using two segments exhibits superior performance over using one segment. The possible reasons are: the two segments can be rationally identified as the same sample as they share a small portion of overlap, and meanwhile they are different enough to increase the task difficulty and thus leads the model to learn a more generalized representation.

\begin{table*}[ht]

  \centering
  \begin{tabular}{l|cc|ccccc|c}
    \toprule
    \textbf{Method}                        & \textbf{Segments} & \textbf{\makecell{length of                                                                                                 \\segment (s)}}   &   \makecell{AS-20K\\mAP (\%)}    & \makecell{SPCV2\\Acc (\%)}   & \makecell{VOX1\\Acc (\%)}
                                           & \makecell{NSYNTH                                                                                                                                \\Acc (\%)} & \makecell{US8K\\Acc (\%)}&\makecell{Average\\Acc (\%)}  \\
    \midrule
    BYOL-A \cite{niizumi_byol_2021}        & single            & 1                           & -             & 92.2          & 40.1          & 74.1          & 79.1          & 71.4          \\
    \midrule
    \textbf{\multirow{4}{*}{ATST-Clip$_{small}$}} & single            & 1                           & 21.0          & 94.3          & 52.3          & 73.8          & 79.3          & 74.9          \\
                                           & two               & 1                           & 19.1          & 91.3          & 50.0          & 74.3          & 76.6          & 73.1          \\
                                           & single            & 6                           & 25.7          & \textbf{94.0} & $57.3 $       & $73.8$        & 80.9          & 76.5          \\
                                           & two               & 6                           & \textbf{27.9} & 93.6          & \textbf{61.9} & \textbf{75.3} & \textbf{82.0} & \textbf{78.2} \\
    \bottomrule
  \end{tabular}
  \caption{Ablation studies on ATST-Clip$_{small}$. Linear evaluation results are shown. "Average" is taken over the last four tasks.}
  \label{tab:segments}
\end{table*}

\begin{figure}[t]
  \centering
  \includegraphics[width=\linewidth]{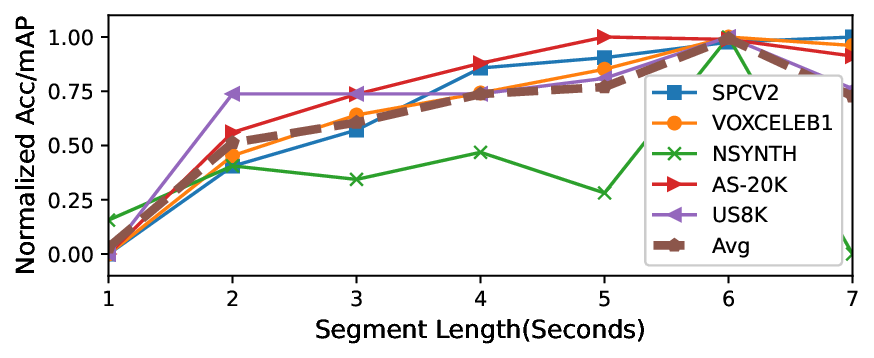}
  \caption{Acc/mAP of ATST-Clip$_{small}$ as a function of segment length, Acc/mAP of each task is normalized into the range of [0,1]. "Avg" denotes averaging over all tasks. }
  \label{fig:length}
\end{figure}

Fig. \ref{fig:length} shows the normalized performance of each task as a function of segment length, where two segments are used. We can observe that as the segment length increases, the performance metrics continue to improve until they reach a maximum at 6 seconds. This further verifies our findings: i) when Transformer encoder is used, increasing the segment length helps to learn more information; ii) when two segments are used, the segment length should be set to make the segments share a proper amount of overlap, and have a proper difficulty for matching them as the same sample.

\subsubsection{Ablations on ATST-Frame}

\begin{table*}[!t]
  \centering
  \begin{threeparttable}

    \begin{tabular}{c|cccc|ccccc|c}
      \toprule
                    & \multicolumn{4}{c|}{} & \multicolumn{5}{c|}{Downstream Tasks}       &                                                                                      \\

      \textbf{Configuration} &  \textbf{\makecell{Augmented\\views}}                                  & \textbf{\makecell{Mask\\teacher}}     & \textbf{\makecell{Mask\\student}}     & \textbf{\makecell{Mask\\strategy}} & \makecell{AS-20K\\mAP (\%)} & \makecell{SPCV2\\Acc (\%)} & \makecell{VOX1\\Acc (\%)} & \makecell{NSYNTH\\Acc (\%)} & \makecell{US8K\\Acc (\%)} & Average \\

      \midrule
      A             &  0 &            & \checkmark & Group         & 8.0    & 63.3  & 25.8 & 56.8   & 60.1     &42.8      \\
      B             &   1                       &            & \checkmark & Random        & 18.0   & 85.2  & 43.7 & 69.9   & 76.1   &  58.6      \\

      C             & 1                            & \checkmark & \checkmark & Group         & 22.5   & 88.2  & 47.0 & 72.9   & 72.9   & 60.7        \\
      D             & 1                           &            & \checkmark & Group         & 28.1   & 92.3  & \bf{67.0} & 72.5   & \bf{84.0} & \bf{68.8}    \\
      E             & 2                           &            & \checkmark & Group         &  \bf{28.5}   & \bf{92.5}  & 59.6 & \bf{74.7}   & 83.4 & 67.7    \\
      \bottomrule
    \end{tabular}
    \caption{Ablation studies on ATST-Frame$_{small}$. Linear evaluation results are shown. }
    \label{tab:abframecom}
  \end{threeparttable}

\end{table*}

\begin{table*}[ht]

  \centering
  \begin{tabular}{l|ccc|ccccc|c}
    \toprule
    \textbf{Method}                        & \textbf{Symmetrical} & \textbf{Loss}  & \textbf{\makecell{Augmented \\branch}}   &   \makecell{AS-20K\\mAP (\%)}    & \makecell{SPCV2\\Acc (\%)}   & \makecell{VOX1\\Acc (\%)}
                                           & \makecell{NSYNTH                                                                                                                                \\Acc (\%)} & \makecell{US8K\\Acc (\%)}&Average  \\
    \midrule
    \textbf{\multirow{3}{*}{ATST-Frame$_{small}$}} & True     & $L_{\theta}+L_{\theta}'$     & Teacher \& Student                          & 28.1          & 92.3          & 67.0          & 72.5          & 84.0          & 68.8          \\
                                           &  False    & $L_{\theta}'$  & Teacher                           & 6.4          & 77.8          &   18.0         & 63.0          & 67.1          & 46.7         \\
                                           & False      &  $L_{\theta}$  &    Student                           & 22.5          & 87.2 &  50.2 &        68.7        & 80.1          & 61.7          \\
    \bottomrule
  \end{tabular}
  \caption{Ablation studies on the symmetrical loss of ATST-Frame$_{small}$.  "Augmented branch" denotes the branch taking as input the augmented view. Linear evaluation results are shown. }
  \label{tab:nonsymmetric}
\end{table*}

\begin{table*}[ht]

  \centering
  \begin{tabular}{l|cc|ccccc|c}
    \toprule
    \textbf{Method}                        & \textbf{Strategy} & \textbf{\makecell{Frequency \\ warping}}   &   \makecell{AS-20K\\mAP (\%)}    & \makecell{SPCV2\\Acc (\%)}   & \makecell{VOX1\\Acc (\%)}
                                           & \makecell{NSYNTH                                                                                                                                \\Acc (\%)} & \makecell{US8K\\Acc (\%)}&Average  \\
    \midrule
    \textbf{\multirow{3}{*}{ATST-Frame$_{small}$}} & Frame-wise         &   True                          & 28.1          & 92.3          & 67.0          & 72.5          & 84.0          & 68.8          \\
                                           &  Frame-wise              &  False                           & 23.5          & 89.4          &  56.9        & 69.0          &   79.7       & 63.7         \\
                                           &  Patch-wise              &  True                           & 16.3          & 77.8          &  35.9        & 71.6          & 75.2          & 55.4         \\
                                           & Patch-wise            & False                           & 23.3          & 80.8 & 48.6 & 70.2                & 79.5          & 60.5         \\
    \bottomrule
  \end{tabular}
  \caption{Ablation studies on comparison of frame-wise and patch-wise strategy. Linear evaluation results are shown. }
  \label{tab:patchwise}
\end{table*}

\begin{table*}[ht]

  \centering
  \begin{tabular}{l|cc|ccccc|c}
    \toprule
    \textbf{Method}                        & \textbf{Symmetrical} & \textbf{\makecell{Augmented \\branch}}   &   \makecell{AS-20K\\mAP (\%)}    & \makecell{SPCV2\\Acc (\%)}   & \makecell{VOX1\\Acc (\%)}
                                           & \makecell{NSYNTH                                                                                                                                \\Acc (\%)} & \makecell{US8K\\Acc (\%)}&Average  \\
    \midrule

    \textbf{ATST-Frame$_{small}$} & True         &   Teacher \& Student                          & 28.1          & 92.3          & 67.0          & 72.5          & 84.0          & 68.8          \\
    \midrule
    \textbf{\multirow{3}{*}{ATST-Frame-data2vec$_{small}^*$}} & False        &   None                      & 24.1          & 92.0          & 58.4         & 73.0         & 81.4          & 65.8          \\
                                           &  False              &  Student                          & 18.6          & 89.3          & 43.7         & 71.5          & 76.0          & 59.8         \\
                                           & True            & Teacher \& Student                           & 21.6          & 90.7 &  52.3 &       70      &  78.4          & 62.6         \\
    \bottomrule
  \end{tabular}
  \caption{Ablation studies on comparison with data2vec-style training target. "ATST-Frame-data2vec$_{small}^*$" denotes ATST-Frame$_{small}$ with data2vec-style\cite{baevski_data2vec_2022} training target. "Augmented branch" denotes the branch takes as input the augmented view. Linear evaluation results are shown. }
  \label{tab:data2vec}
\end{table*}

Table \ref{tab:abframecom} shows the ablation results on the effectiveness of ATST-Frame components. 

\textbf{Data augmentation:} Compared with the no augmentation case (configuration A in Table \ref{tab:abframecom}), using data augmentation (configuration B, C, D, E) brings a significant performance improvement on all tasks. This means data augmentation is able to properly increase the task difficulty and to encourage the model to learn more meaningful audio representations. Augmenting two views (for both teacher and student branches, configuration E) leads to a large task difficulty, and achieves better results on three out of five tasks. Only augmenting one view (for either student or teacher branch, configuration D) achieves slightly worse performance than augmenting two views on AS-20K and SPCV2, but much better performance on VOX1, thus has a better average result. This is consistent with our observations in the ablation studies of ATST-Clip that better performance can be achieved with a balanced difficulty of the pre-training task. 

\textbf{Masking:}
In configuration C, both the student and teacher branches are masked with the same time index, thus the two branches need to predict the masked frames from unmasked frames, and the predictions should be matched.
In configuration D, only the student branch is masked, while the teacher branch sees the whole audio clip. We can see that masking both branches performs worse than masking only the student branch. The possible reasons are i) the teacher branch provides more meaningful guidance for the student branch when seeing the frames that are not visible to the student branch; ii) the teacher encoder consistently sees unmasked input for pre-training and downstream tasks. As for the masking strategy, group masking that forces N adjacent frames to be masked together performs better than random masking \addnote[ConfigurationB]{1}{(Configuration B)}. This is consistent with the observations in the speech pre-training works\cite{baevski_wav2vec_2020,baevski_data2vec_2022}.

Based on the above analysis, the proposed ATST-Frame is set up with configuration D. Unless noted, the following experiments of ATST-Frame use configuration D by default. 

\addnote[symmetricloss]{1}{\textbf{The symmetrical loss:} In ATST-Frame, augmentation is applied to one of the two views. As the symmetrical loss is used, both the teacher branch and the student branch see the augmented view during training. We conduct experiments by using only $L_{\theta}$ or $L'_{\theta}$ to evaluate which one of teacher and student branches is more important to see the augmented view. 
The results are shown in Table \ref{tab:nonsymmetric}, which shows that it is more important for the student branch than the teacher branch to see the augmented view, and using the symmetrical loss outperforms the case that only one branch sees the augmented view. }

\addnote[patch-wise]{1}{\textbf{Patch-wise strategy and frequency warping:} ATST-Frame uses frame-wise strategy for log-mel spectrogram, while other works\cite{gong_ssast_2022,baade_mae-ast_2022} have reported that patch-wise strategy exhibits better performance than frame-wise strategy on sound event/scene classification task, as sound events/scenes have complex frequency
structure, which can be better captured by the frequency split
of patch-wise models \cite{gong_ssast_2022}.
To testify the frame-wise strategy of our ATST-Frame model, we further conduct experiments using patch-wise strategy in the framework of ATST-Frame. Specifically, we organize the log-mel spectrogram into patches in the size of 16 frequency bins $\times$ 16 frames, which leads to the same number of tokens as ATST-Frame for a 64-bin log-mel spectrogram. For patch-wise models, frequency warping conflicts with the principle of the patch-wise loss, as it distorts the patch correspondences. Therefore, we conducted experiments both with or without using frequency warping. Note that Mixup is always used. The results are shown in \ref{tab:patchwise}. Without using frequency warping, the frame-wise model noticeably performs better than the patch-wise model on speech tasks (SPCV2 and VOX1), which is consistent with the observations in other works\cite{gong_ssast_2022,baade_mae-ast_2022}. However, we do not observe the advantage of patch-wise strategy on sound event/scene classification (AS-20K and US8K), where patch-wise strategy and frame-wise strategy are comparable. The frame-wise model significantly benefits from frequency warping for all tasks whereas the patch-wise model does not. Frequency warping (FW) encourages learning FW-invariant representations, which may help to learn the spectral pattern, even for the complex spectral structure of sound events/scenes. 
 Overall, within the framework of ATST-Frame, the frame-wise strategy is suitable for both speech and sound events/scenes, and frequency warping helps to largely improve the performance. 
}

\addnote[data2vec]{1}{\textbf{Using the training target of data2vec:} 
We apply the training target of data2vec \cite{baevski_data2vec_2022} to our ATST-Frame model (referred to as ATST-Frame-data2vec). Specifically, the last 8 blocks of the teacher encoder are averaged to form the training target; the projectors are removed; the predictor is replaced with a linear projection. Although the original data2vec does not use data augmentation and symmetrical loss, 
we also test how will data augmentation and symmetrical loss perform when used to ATST-Frame-data2vec. The results are reported in Table \ref{tab:data2vec}. It can be seen that the data2vec target performs well, but it does not benefit from data augmentation. }

\subsection{Results on Clip-level Downstream Tasks}

\subsubsection{Linear Evaluation Results}

Table \ref{tab:le_result} shows the linear evaluation results on six tasks.
For a fair comparison, we compare with other methods that also use Audioset for pre-training and have also reported the linear evaluation results in their papers, including TRILL \cite{shor2020towards}, COLA \cite{saeed_contrastive_2020}, BYOL-A \cite{niizumi_byol_2021}, BYOL-A-v2\cite{niizumi_byol_2023}, SF NFNET-F0\cite{wang_towards_2022} and M2D\cite{niizumi_masked_2023}. 
The proposed ATST-Clip is developed based on BYOL-A and BYOL-A-V2, using a Transformer encoder and a new view creation strategy. It can be seen that ATST-Clip noticeably outperforms BYOL-A and BYOL-A-V2 on all tasks, which indicates that our modifications are very effective. On average, the proposed models and the recently proposed M2D model perform better than other models. The performance of the proposed models and M2D are comparable, as M2D performs better on SPCV2, NSYNTH and US8K with small advantages, while the proposed ATST-Frame performs better on VOX1. Among the two proposed models, ATST-Clip outperforms ATST-Frame for all the tasks except for VOX1. ATST-Clip is dedicated to learning clip-level representation, its embedding is more representative for the audio clip than the one obtained by averaging the frame-level embeddings of ATST-Frame. However, the drawbacks of ATST-Frame are not significant., which means the average of its frame-level embeddings is still an effective clip-level representation. 


\def\a{true}
\def\b{false}
\setlength\tabcolsep{3pt}
\begin{table}[t]

  \centering

  \footnotesize
  \scalebox{0.85}{
    \begin{threeparttable}
      \begin{tabular}{l|cccccc}
        \toprule
        \textbf{Method}                    & \makecell{AS-20K                                          \\mAP (\%)}   & \makecell{SPCV2\\Acc (\%)}    & \makecell{VOX1\\Acc (\%)}
                                           & \makecell{NSYNTH                                          \\Acc (\%)} & \makecell{US8K\\Acc (\%)} & \makecell{FSD50K \\ mAP (\%)} \\
        \midrule
        TRILL \cite{shor2020towards}       & -                & -      & 17.9   & -      & -   & -        \\
        COLA \cite{saeed_contrastive_2020} & -                & $62.4$ & $29.9$ & $63.4$ & -     & -      \\
        BYOL-A \cite{niizumi_byol_2021}    & -                & $92.2$ & $40.1$ & $74.1$ & 79.1        \\
        BYOL-A-V2 \cite{niizumi_byol_2023} & -                & 93.1   & 57.6   & 73.1   & 79.7 & 44.8 \\
        SF NFNet-F0\cite{wang_towards_2022} & - & 93.0 & 64.9 & \bf{78.2} & - \\
        M2D\cite{niizumi_masked_2023} &- &\textbf{95.4} & 73.1 & 76.9 & \textbf{87.6} & - \\
        \midrule
        {ATST-Clip (ours)}          & \bf{33.8}             & 95.1   & 72.0   & 76.2   & 85.8 & \bf{58.5}
        \\


        {ATST-Frame (ours)}         & 33.0             & 94.9   & \bf{77.4}   & 75.9   & 85.8 & 55.1
        \\


        \bottomrule
      \end{tabular}

    \end{threeparttable}
  }

  \caption{ Linear evaluation results on clip-level downstream tasks. The scores of comparison models are quoted from their papers. }
  \label{tab:le_result}
\end{table}

\begin{table*}[ht]

  \centering
  \footnotesize
  \scalebox{1.0}{
    \begin{threeparttable}
      \begin{tabular}{lcccccccc}
        \toprule
        \textbf{Method}                                & \# Param  & \makecell{Pre-training\\data} & \makecell{AS-2M                                 \\ mAP (\%)} & \makecell{AS-20K\\mAP (\%)}    & \makecell{SPCV2\\Acc (\%)}   & \makecell{VOX1\\Acc (\%)} & \makecell{FSD50K \\ mAP (\%)} & \makecell{NSYNTH \\ Acc (\%)} 
        \\
        \midrule
        \multicolumn{5}{l}{\textbf{Supervised Methods} }                                                             \\
        PANN \cite{kong2020panns}                       & 81M &         & 43.9            & 27.8 & -      & -      & -  & -  \\

        PSLA \cite{pascual_learning_2019}                &14M &         & 44.4            & 31.9 & -      & -      & 55.4 & - \\

        AST \cite{gong_ast_2021}                         & 86M&         & 45.9            & 34.7 & 98.1   & -      & -   & - \\
        HTS-AT \cite{chen_hts-at_2022}  & 31M & & 47.1 & - & 98.0 &  - & - & - \\
        PassT \cite{koutini_efficient_2022} & 86M& & 47.1 & - & -& - & 65.3 & -  \\
        KD-AST \cite{gong_cmkd_2022}         &  86M          &         & 47.1            & -    & -      & -      & 62.9 & - \\
        \midrule[0.2pt]
        \multicolumn{5}{l}{\textbf{Self-supervised Methods} }                                                        \\

        SSAST-PATCH\cite{gong_ssast_2022} & 89M & AS+LS & - & 31.0 & 98.0 & 64.2 & - & -  \\
        SSAST-FRAME\cite{gong_ast_2021} &89M & AS+LS &- & 29.2 &98.1 & 80.8 & - & - \\
        Conformer-Based\cite{srivastava_conformer-based_2022} &88M & 67K hours *        & 41.5               & 27.6 & -      & -   &-  & -       \\

        MAE-AST-PATCH\cite{baade_mae-ast_2022}    &86M & AS+LS   & -               & 30.6 & 97.9   & -   & -  & -    \\
        MAE-AST-FRAME\cite{baade_mae-ast_2022}    &86M & AS+LS   & -               & 23.0 & 98.0   & 63.3   & -  & -   \\
        ASiT\cite{atito_asit_2022}                      &85M & AS      & -               & 35.2 & \textbf{98.8}   & 63.1  & -   & -     \\
        data2vec\cite{baevski_data2vec_2022} & 94M & AS & -&34.5&-&-&- & -\\
        MaskSpec\cite{10095691}  & 86M & AS & 47.1 & 34.7 & 97.6 & - & - & -  \\
        MSM-MAE\cite{niizumi_masked_nodate} $^{\dagger}$ & 86M & AS & - & 36.7 & 98.4 & 95.3 &  -  & - \\

        Audio-MAE (local) \cite{huang_masked_2023} &86M & AS      & 47.3            & 37.0 & 98.3   & 94.8   & - & -     \\
        BEATs$_{iter3}$ \cite{chen_beats_2022}          & 90M& AS      & 48.0            & 38.3 & 98.3   & -      & -   & -  \\        
       
                       BEATs$_{iter3+}$ \cite{chen_beats_2022} **          & 90M& AS      & 48.6            & 38.9 & 98.1  & -      & -   & -   \\
        M2D\cite{niizumi_masked_2023}                   & 86M & AS      & -               & 37.4 & 98.5   & 94.4   & -  & -  \\

        \midrule

        \multicolumn{5}{l}{\textbf{Ours} } \\                              
        {ATST-Clip }              & 86M         & AS      & 45.2            & 37.9 & 98.0   & 95.5   & 63.4 & 78.6 
        \\
        \vspace{1mm}
        {ATST-Frame }             &   86M       & AS      & 48.0            &39.0 & 98.1   & 97.3   & 61.8 & 79.2 \\


         {ATST-C2F} ** &86M & AS & \bf{49.7} & \bf{40.5} & 98.4& \bf{97.5} & \bf{65.5} & 79.2  \\
         
        \textcolor{mygray}{ATST-F2C **}  &\textcolor{mygray}{86M} & \textcolor{mygray}{AS} & \textcolor{mygray}{46.8} & \textcolor{mygray}{39.0} & \textcolor{mygray}{98.1}& \textcolor{mygray}{95.5} & \textcolor{mygray}{64.6} & \textcolor{mygray}{\bf{79.8}}  \\

        \bottomrule
      \end{tabular}
      \begin{tablenotes}
      \item{*} Self-hold dataset\cite{srivastava_conformer-based_2022}.
      \item{$\dagger$} Results are quoted from M2D\cite{niizumi_masked_2023}.
      \item{**} Perform knowledge distillation across two models at the finetuning stage.
      \end{tablenotes}

    \end{threeparttable}
  }
  \caption{ Finetuning results on clip-level downstream tasks. The scores of comparison models are quoted from their papers. AS and LS denote AudioSet and Librispeech\cite{panayotov2015librispeech}, respectively.  }
  \label{tab:finetune_result}
\end{table*}

\subsubsection{Fine-tuning Results}

Linear evaluation cannot fully reflect the capabilities of pre-trained models, as normally the models can be further fine-tuned with the data  of downstream tasks. Fine-tuning experiments are conducted on the tasks of multi-label audio event classification (AS-2M, AS-20K and FSD50K), Spoken command recognition (SPCV2), speaker identification (VOX1) and musical instrument classification (NSYNTH). We compare with two groups of prior methods: supervised methods and self-supervised methods. The results are shown in Table \ref{tab:finetune_result}. 

\textbf{ATST-Frame outperforms ATST-Clip.} After fine-tuning, the performance of ATST-Frame is better than ATST-Clip on five out of six tasks. As mentioned above, ATST-Frame does not explicitly learn clip-level representation during pre-training. However, fine-tuning allows the adjustment of the pre-trained parameters to fit a specific downstream task. 
ATST-Frame is pre-trained by maximizing the agreement of frame-level embeddings, which is more fine-grained and challenging compared with ATST-Clip. This may help ATST-Frame to learn more sophisticated knowledge and network parameters (such as the self-attention parameters), which happen to be a better initial setting for fine-tuning even on clip-level downstream tasks.

\textbf{Comparison with supervised methods.} The proposed ATST-Frame outperforms the supervised methods on AS-2M, AS-20K and SPCV2. The proposed ATST-C2F model further improves the performance, and outperforms the supervised methods on all tasks. This is encouraging for the field of audio self-supervised learning, as we no longer need to annotate audio data for pre-training when we want to further scale up the dataset. Compared with supervised pre-training, self-supervised pre-training does not suffer from the problem of inaccurate and erroneous labels.

\textbf{Comparison with other self-supervised methods.} Compared with other self-supervised methods, the proposed ATST-Frame achieves comparable or better performance on all tasks. In particular, compared with the recent state-of-the-art self-supervised method BEATs$_{iter3+}$\cite{chen_beats_2022}, ATST-Frame achieves the same performance on AS-2M, and better performance on AS-20K. This indicates that ATST-Frame is more effective with less fine-tuning data than BEATS. 

\textbf{Combination through knowledge distillation.} 
ATST-Clip and ATST-Frame learn complementary features in the pre-training stage. Combining ATST-Clip and ATST-Frame through knowledge distillation can further improve the performance, as shown by the results of ATST-C2F in Table \ref{tab:finetune_result}. Even though ATST-Clip performs worse than ATST-Frame on AS-2M and AS-20K, as a teacher, it still successfully helps to fine-tune ATST-Frame to achieve better performance. 
Performing knowledge distillation the other way around, i.e. from ATST-Frame to ATST-Clip, does not perform well on most of the tasks. 
\addnote[sophisticated]{1}{
This verifies that ATST-Frame conveys more fine-grained and semantically complicated information than ATST-Clip
}, and ATST-Frame should be used as the final model.
BEATs$_{iter3+}$\cite{chen_beats_2022} also performs knowledge distillation across models at the fine-tuning stage, specifically, it uses the fine-tuned BEATs$_{iter2}$ model as a teacher to fine-tune the final BEATs$_{iter3}$ model. Thence, BEATs$_{iter3+}$ can be regarded as a fair comparison with the proposed ATST-C2F model.

\subsection{Results on Frame-level Downstream Task - DESED}
\label{sec:dcase}

\begin{table}[h]
  \centering
  \footnotesize
  \scalebox{1}{
    \begin{threeparttable}
        \begin{tabular}{lcccc}
        \toprule 
        \textbf{Method} & $\tau$ (ms) & \textbf{learning rate} & \textbf{$\text{PSDS}_1$} & \textbf{$\text{PSDS}_2$}\\
        \midrule
        \multicolumn{5}{l}{\textbf{Linear Evaluation}} \\
        BYOL-A-V2-40ms \cite{niizumi_byol_2023}      & 40 & 0.01 & 0.024 & 0.181  \\
        SSAST-FRAME-20ms \cite{gong_ssast_2022}      & 20 & 0.1 & 0.028 & 0.166 \\
        SSAST-FRAME-40ms \cite{gong_ssast_2022}      & 40 & 0.1 & 0.096 & 0.266 \\
        SSAST-PATCH     \cite{gong_ssast_2022}       & 160 & 0.1 & 0.179 & 0.315  \\
        MAE-AST-FRAME-20ms \cite{baade_mae-ast_2022} & 20 & 0.1 & 0.031 & 0.234 \\
        MAE-AST-FRAME-40ms \cite{baade_mae-ast_2022} & 40 & 0.1 & 0.081 & 0.293 \\
        MAE-AST-PATCH   \cite{baade_mae-ast_2022}    & 160 & 0.1 & 0.225 & 0.442 \\
        Audio-MAE (local) \cite{huang_masked_2023}   & 160 & 0.1 & 0.218 & 0.401 \\
        BEATs$_{iter3}$ \cite{chen_beats_2022}       & 160 & 0.1 & 0.177 & 0.358 \\
        M2D \cite{niizumi_masked_2023}               & 160 & 0.1 & 0.234 & 0.438 \\
        \midrule
        \multicolumn{5}{l}{\textbf{Ours}} \\
        ATST-Clip                            & 40 & 0.1 & 0.115 & 0.293   \\
        ATST-Frame                           & 40 & 0.1 & \textbf{0.304} & \textbf{0.507} \\
        \midrule
        \midrule
        \multicolumn{5}{l}{\textbf{Finetuning}} \\
        BYOL-A-V2-40ms \cite{niizumi_byol_2023}      & 40 & 0.01 & 0.030 & 0.219 \\
        SSAST-FRAME-20ms \cite{gong_ssast_2022}      & 20 & 0.05 & 0.046 & 0.235 \\
        SSAST-FRAME-40ms \cite{gong_ssast_2022}      & 40 & 0.1 & 0.132 & 0.325 \\
        SSAST-PATCH     \cite{gong_ssast_2022}       & 160 & 0.1 & 0.236 & 0.459 \\
        MAE-AST-FRAME-20ms \cite{baade_mae-ast_2022} & 20 & 0.05 & 0.123 & 0.346 \\
        MAE-AST-FRAME-40ms \cite{baade_mae-ast_2022} & 40 & 0.1 & 0.235 & 0.418 \\
        MAE-AST-PATCH   \cite{baade_mae-ast_2022}    & 160 & 0.05 & 0.281 & 0.573 \\
        Audio-MAE (local) \cite{huang_masked_2023}   & 160 & 0.1 & 0.254 & 0.509 \\
        BEATs$_{iter3}$ \cite{chen_beats_2022}       & 160 & 0.1 & 0.282 & 0.584 \\
        M2D \cite{niizumi_masked_2023}               & 160 & 0.1 & 0.267 & 0.500 \\
        \midrule
        \multicolumn{5}{l}{\textbf{Ours}} \\
        ATST-Clip                          & 40 & 0.05 & 0.223 & 0.422      \\
        ATST-Frame                         & 40 & 0.01 & \textbf{0.361} & 0.581 \\
        ATST-C2F                         & 40 & 0.1 & 0.357 & \textbf{0.607}  \\
        \textcolor{mygray}{ATST-F2C}                         & \textcolor{mygray}{40} & \textcolor{mygray}{0.1} & \textcolor{mygray}{0.259} & \textcolor{mygray}{0.445} \\
        \bottomrule
        \end{tabular}     
    \end{threeparttable}
  }
  \caption{Results on the frame-level downstream task, DESED.  {$\tau$} stands for temporal resolution.}
  \label{tab:linear_SED}
\end{table}

\subsubsection{Comparison Methods}

We compare with six SSL pre-trained models: BYOL-A-v2 \cite{niizumi_byol_2023}, SSAST \cite{gong_ssast_2022}, MAE-AST \cite{baade_mae-ast_2022}, Audio-MAE \cite{huang_masked_2023}, BEATs \cite{chen_beats_2022} and M2D \cite{niizumi_masked_2023}. 
Sound event detection requires to perform frame-level multi-class classification. As mentioned in Section \ref{sec:sedsetup}, the proposed models can be directly used for this task by adding a linear classifier on top of their frame-level representations, with a temporal resolution of 40 ms per frame. The comparison models are pre-trained either frame-wisely or patch-wisely. The frame-wise models, e.g. SSAST and MAE-AST, can also be directly used for this task, with a temporal resolution of 20 ms per frame. According to the setup of the proposed models, we also evaluate SSAST and MAE-AST with a temporal resolution of 40 ms per frame, by applying average pooling to the frame-level representations.  
As for the patch-wise models, we average the patch-level representations for each time interval to obtain the interval/frame-level representations, except for M2D, since its authors propose to concatenate instead of average the patch-level representations \cite{niizumi_masked_2023}. 
Note that, the patch-wise models have a coarser temporal resolution, i.e. 160 ms per frame.

All the pre-trained models are fine-tuned by ourselves using the SED supervised dataset. For linear evaluation, pre-trained models are frozen, and only the two dense layers of the linear classifier are trained. In fine-tuning experiments, for all the Transformer-based models, we unfreeze the last Transformer block. For BYOL-A-v2, we unfreeze the entire model for fair comparison such that the amount of the trainable parameters of each model are similar. The best learning rate for each model has been carefully searched, which is also given in Table \ref{tab:linear_SED}.

\subsubsection{Results Analysis}

\addnote[DESEDTraining2]{1}{Table \ref{tab:linear_SED} shows the results. These results are deviated from the results reported in the DCASE challenge, that is because of the different experimental setups as discussed in Sec.~\ref{sec:DESEDTraining}. The objective of this study is to conduct fair  comparison between different SSL models, instead of pursuing the SOTA performances on this dataset.} It can be seen that, as expected, relative to linear evaluation, the performance of all models can be improved by fine-tuning with the SED dataset. As the fine-tuning setup is more practically important than linear evaluation, we mainly analyze the fine-tuning results in the following, and most of the analyses are valid for the linear evaluation results as well. 

BYOL-A-V2 does not achieve reasonable performance, possibly due to its limited capacity for sequential processing with a two-layer CNN architecture.
For SSAST \cite{gong_ssast_2022} and MAE-AST \cite{baade_mae-ast_2022}, increasing the temporal resolution of their frame-wise models from 20 ms to 40 ms largely improves the performance. The possible reasons are that the frame-level representations get more stable when averaging two frames, and meanwhile, the 40 ms temporal resolution is still fine enough for tracking the time variation of sound events. This could also be because the characteristics of one event cannot be well represented without sufficiently long frames. However, the performances of their frame-wise models still largely lag behind their patch-wise models. This is consistent with the observations in \cite{gong_ssast_2022,baade_mae-ast_2022} that, the patch-wise models are more suitable for sound events, while the frame-wise models are more suitable for speech signals. Sound events have more complex frequency structure, which can be better captured by the frequency split of patch-wise models. Among the comparison models, BEATs performs the best in terms of both $\text{PSDS}_1$ and $\text{PSDS}_2$, and MAE-AST-PATCH achieves close performance with BEATs. 

The proposed ATST-Clip does not work well, as it is trained for learning global representation, which does not automatically lead to good frame-level representations. By leveraging the proposed frame-level training criterion and thus learning better frame-level representations, ATST-Frame largely improves the performance over ATST-Clip.  Compared with the best comparison model, i.e. BEATs, ATST-Frame achieves much better $\text{PSDS}_1$, and similar $\text{PSDS}_2$. 
$\text{PSDS}_1$ emphasizes the time localization accuracy of sound events, thence the better $\text{PSDS}_1$ of ATST-Frame means a better temporal detection performance, which is possibly due to the finer temporal resolution of ATST-Frame compared with BEATs, i.e. 40 ms versus 160 ms. 
$\text{PSDS}_2$ emphasizes the recognition accuracy of sound events. The similar $\text{PSDS}_2$ of ATST-Frame and BEATs reflect the similar representation quality of them. This is consistent with the results on the clip-level AS-2M task, ATST-Frame also performs similarly with BEATs as shown in Table \ref{tab:le_result}. It is important to note that, the good performance of ATST-Frame conflicts with the observations in \cite{gong_ssast_2022,baade_mae-ast_2022} that patch-wise models are more suitable for sound events than frame-wise models. As discussed in the ablation study, the success of ATST-Frame is possibly attributed to the frequency warping operation, which helps to capture the complex frequency structure of sound events.

Knowledge distillation is also applied to combine ATST-Clip and ATST-Frame. The results of ATST-C2F show that, taking ATST-Clip as a teacher for fine-tuning ATST-Frame, $\text{PSDS}_2$ can be further improved, while $\text{PSDS}_1$ is slightly decreased. This means the knowledge learned by ATST-Clip  is still complementary for improving the accuracy of frame-level representations, but will slightly blur the time localization.    


\begin{table}[tb]
  \centering
  \footnotesize
  \scalebox{1}{
      \begin{tabular}{l|c|cc|cc}
        \toprule 
        \multirow{3}{*}[-0.5em]{\textbf{Method}} & 
        \multirow{3}{*}[-0.3em]{\textbf{\makecell{Learning\\ rate}}} & 
        \multicolumn{2}{c|}{\textbf{$\text{PSDS}_{1} $}} &
        \multicolumn{2}{c}{\textbf{$\text{PSDS}_{2} $}} \\
        \cmidrule{3-6}
        &
        & w/o  & with
        & w/o  & with \\          
        &
        & var-pen & var-pen
        & var-pen & var-pen \\
        \midrule 
        \multicolumn{3}{l}{\textbf{Linear Evaluation}} \\
        BYOL-A-V2-40ms \cite{niizumi_byol_2023}      & 0.5 & 0.087 & 0.0 & 0.083 & 0.0 \\
        SSAST-PATCH     \cite{gong_ssast_2022}       & 0.5 & 0.048 & 0.0 & 0.067 & 0.0 \\
        MAE-AST-PATCH   \cite{baade_mae-ast_2022}    & 0.5 & 0.116 & 0.0 & 0.185 & 0.0 \\
        Audio-MAE (local) \cite{huang_masked_2023}   & 0.5 & 0.073 & 0.0 & 0.107 & 0.0 \\
        BEATs$_{iter3}$ \cite{chen_beats_2022}       & 0.5 & 0.034 & 0.0 & 0.062 & 0.0 \\
        M2D \cite{niizumi_masked_2023}               & 0.5 & 0.182 & 0.0 & 0.301 & 0.039 \\
        \midrule
        \multicolumn{6}{l}{\textbf{Ours}} \\
        ATST-Clip  & 0.5 & 0.120 & 0.0 & 0.201 & 0.001 \\
        ATST-Frame & 0.5 & \textbf{0.207} & \textbf{0.008} & \textbf{0.304} & \textbf{0.048} \\
        \midrule
        \midrule
        \multicolumn{6}{l}{\textbf{Finetuning}} \\
        BYOL-A-V2-40ms \cite{niizumi_byol_2023}      & 0.5 & 0.110 & 0.0 & 0.243 & 0.027 \\
        SSAST-PATCH     \cite{gong_ssast_2022}       & 0.5 & 0.243 & 0.017 & 0.411 & 0.122 \\
        MAE-AST-PATCH   \cite{baade_mae-ast_2022}    & 0.5 & 0.274 & 0.039 & 0.481 & 0.187 \\
        Audio-MAE (local) \cite{huang_masked_2023}   & 0.5 & 0.276 & 0.038 & 0.476 & 0.182 \\
        BEATs$_{iter3}$ \cite{chen_beats_2022}       & 0.5 & 0.290 & 0.045 & 0.491 & 0.186 \\
        M2D \cite{niizumi_masked_2023}               & 0.1 & 0.292 & 0.042 & 0.509 & 0.199 \\
        \midrule
        \multicolumn{6}{l}{\textbf{Ours}} \\
        ATST-Clip  & 0.5 & 0.328 & 0.083 & 0.478 & 0.178 \\
        ATST-Frame & 0.5 & 0.347 & 0.069 & 0.538 & 0.152 \\
        ATST-C2F   & 0.5 & \textbf{0.374} & \textbf{0.125} & \textbf{0.572} & \textbf{0.266} \\
        \textcolor{mygray}{ATST-F2C} & \textcolor{mygray}{0.5} & \textcolor{mygray}{0.323} & \textcolor{mygray}{0.075} & \textcolor{mygray}{0.470} & \textcolor{mygray}{0.163} \\
        \bottomrule
        \end{tabular}
    }
  \caption{Results on the frame-level downstream task, SED of the strongly-labeled AudioSet. `var-pen' stands for the performance variance penalty term.   } 
  \label{tab:audioset_strong}
\end{table}

\subsection{Results on Frame-level Downstream Task - Strongly-labeled AudioSet}
\label{sec:audioset_strong}

Table \ref{tab:audioset_strong} shows the results on the strongly-labeled AudioSet. According to the DESED performances, only the best-performing model for each comparison method is evaluated. 
Considering the large data size of strongly-labeled AudioSet, we only search over 3 different learning rates for each model, i.e. 0.05, 0.1 and 0.5. 
As mentioned in Sec.~\ref{sec:frame_metric}, we evaluate the models by the PSDSs with or without applying the performance variance penalty term. 

For all the models, the scores with variance penalty are much lower than the ones without variance penalty, which means the performance variance across classes for all models are very large. The scores with variance penalty for linear evaluation could be reduced to 0 for most of the models. This reflects the data imbalance and task difficulty of the strongly-labeled AudioSet. 

After finetuning, The BYOL-A-v2 model has a large performance gap comparing with other Transformer-based models. 
With better learned frame-level representations, the proposed ATST-Frame model has an obvious advantage over the comparison models and ATST-Clip, when variance penalty is not applied. However, ATST-Clip has a better stability of performance across classes, and thus outperforms ATST-Frame when applying variance penalty. When combining ATST-Frame and ATST-Clip, the performance measures are largely improved by ATST-C2F, and the model is improved in terms of both classification accuracy and performance stability.  


\subsection{Results on HEAR benchmark} We also evaluate the proposed models on the HEAR benchmark \cite{turian_hear_2022}, which includes 17 clip-level and 2 frame-level tasks. We successfully downloaded 18 tasks. The Hear benchmark trains a shallow MLP classifier on top of frozen embeddings. We use the official hear-eval-kit \footnote{https://github.com/hearbenchmark/hear-eval-kit}, and our embeddings are extracted in the same way as we did in our linear evaluation experiments except that for frame-level tasks, we concatenate outputs of all the blocks. 
Table \ref{tab:hear} shows the results. As a baseline, we quote the best result for each task from the HEAR leaderboard \footnote{https://hearbenchmark.com/hear-leaderboard.html}, denoted as `Best' in the table. \addnote[hear]{1}{It is worth noting that there are two frame-level tasks, i.e. DCASE 2016 and Maestro 5h. On DCASE 2016, both ATST-Clip and ATST-Frame perform better than the best baseline. However, the best baseline performs much better than the proposed models for Maestro 5h.  Maestro 5h is a piano music transcription task, aiming to extract pitch and onset from raw audio. The data augmentation of RRC and frequency warping in ATST encourage the model to learn frequency-changing-invariant representations, which may be not suitable for pitch learning, as pitch is sensitive to frequency changing. 
This phenomenon is also observed on the clip-level pitch estimation task, i.e. NSynth Pitch 5h.} Overall, both the proposed ATST-Clip and ATST-Frame achieve better performance than the best baseline on  five tasks. This is remarkable considering the fact that the best baseline results quoted here for different tasks are achieved by 14 different submissions. Moreover, some of the best baseline results are obtained by the model especially trained for the specific tasks, as HEAR benchmark does not limit the pre-training methods (supervised or unsupervised) and pre-training datasets. 

\begin{table}[!t]
  \centering
  \begin{threeparttable}

    \begin{tabular}{l|ccc}
      \toprule
                         & ATST-Clip & ATST-Frame & Best \\ 
      \midrule
      Beehive                    & 58.3       &    64.6  &  87.8  \\ 
      Beijing Opera              & 95.3          & 95.8   & 97.5 \\ 
      CREMA-D                   & \underline{76.0}            &\underline{76.7} & 75.2  \\ 
      DCASE 2016 *               & \underline{93.7}             &\underline{95.7}  & 92.5 \\
      ESC-50                    & 91.2        & 89.0   & 96.1 \\  
      FSD50K                     & 59.5         & 55.7 & 64.1 \\ 
      Gunshot                    & \underline{98.8}         &94.3  & 96.7  \\ 
      GTZAN Genre                & 87.7           & 88.3  & 90.8 \\ 
      GTZAN Music/Speech            & 99.2          &\underline{ 100.0} & 99.2\\ 
      Libricountl                & 78.2         & 78.1  & 78.5 \\ 
      Maestro 5h *                & 18.9         &24.4 & 46.9 \\ 
      Mridangam Stroke           & \underline{97.7}            & 97.5 & 97.5\\ 
      Mridangam Tonic            & \underline{96.7}           & \underline{96.9}  & 94.1\\ 
      NSynth Pitch 5h            & 67.8         &68.6  & 87.8 \\ 
      Speech command 5h        & 93.1          & 92.6 & 97.6 \\ 
      Speech command full         & 95.5         &95.1  & 97.8 \\

      Vocal Imitation           & 18.5           & \underline{22.3} & 21.5 \\ 
      VoxLingua107 top 10          & 53.9            & 66.9 & 72.2 \\ 

      \bottomrule
    \end{tabular}
      \begin{tablenotes}
      \item{*} frame-level task
      \end{tablenotes}
  \end{threeparttable}
  \caption{Results on the HEAR benchmark. `Best' denotes the best result in the HEAR leaderboard. Underlined scores denote better performance than `Best'. }
  \label{tab:hear}
\end{table}

\section{Conclusion}
\addnote[conclusion]{1}{
In this paper, based on the teacher-student scheme of BYOL, we have proposed two effective self-supervised audio pre-training methods, ATST-Clip and ATST-Frame, specifically crafted to learn clip-level and frame-level representations, respectively, enabling effective audio understanding.\\
\indent The proposed methods have been extensively evaluated on a variety of downstream tasks, including seven clip-level tasks and two frame-level tasks, covering multiple audio domains: environmental sound, speech and music. Both ATST-Clip and ATST-Frame demonstrated their outstanding  capabilities of learning audio representations compared with previous state-of-the-art methods. Furthermore, ATST-Clip offers complementary knowledge to ATST-Frame, and these knowledge can be effectively distilled to  ATST-Frame at the fine-tuning stage. Especially, the proposed methods achieve new SOTA scores on the AudioSet-2M and AudioSet-20K datasets, with the precision of 49.7\% and 40.5\% (without model ensembling), respectively.
Furthermore, this work also provides a new benchmark for applying pre-trained models to frame-level downstream tasks, on two sound event detection datasets. The frame-level downstream tasks have rarely been studied in the field, and hopefully this work would fill this gap.  \\
\indent We have open-sourced our code online for the research community to replicate and expedite future research. As the scope of this study is limited to audio classification tasks, future work may extend our models to audio generation tasks. }
 

\bibliographystyle{IEEEtran}

\bibliography{mybib}

 


\vfill

\end{document}